\documentclass[a4paper,11pt]{article}
\usepackage{jcappub} 
\usepackage{lineno}
\usepackage{hanging}
\usepackage{orcidlink}

\usepackage{graphicx}
\usepackage[sort&compress,numbers]{natbib}
\usepackage{float}
\usepackage{xcolor}

\newcommand{\hMsun}{h^{-1}\mathrm{M_\odot}}

\newcommand{\hkpc}{h^{-1}\mathrm{kpc}}
\newcommand{\hMpc}{h^{-1}\mathrm{Mpc}}
\newcommand{\hGpc}{h^{-1}\mathrm{Gpc}}
\newcommand{\LambdaCDM}{\Lambda \rm{CDM}}

\newcommand{\kms}{\mathrm{km\,s^{-1}}}

\newcommand{\Uchuu}{\textsc{Uchuu }}
\newcommand{\Uchuunospace}{\textsc{Uchuu}}
\newcommand{\vpeak}{$V_\mathrm{peak}$ }

\newcommand{\edited}[1]{\textcolor{black}{#1}}

\newcommand{\minScale}{1 }
\newcommand{\maxScale}{100 }
\newcommand{\minFit}{3 }
\newcommand{\maxFit}{70 }
\newcommand{\minBias}{10 }
\newcommand{\maxBias}{80 }
\title{\boldmath DESI DR2 Reference Mocks: Clustering results from \textsc{UCHUU} ELGs and QSOs}

\usepackage{hyperref}

\usepackage{orcidlink}

\author[a]{{R.~Vaisakh}\orcidlink{0009-0001-2732-8431},}
\author[b]{{J.~Lasker}\orcidlink{0000-0003-2999-4873},}
\author[a]{{R.~Kehoe},}
\author[a]{{A.~Amalbert},}
\author[a]{{N.~Khan},}
\author[c]{{E.~Fernández-García}\orcidlink{0009-0006-2125-9590},}
\author[c]{{F.~Prada}\orcidlink{0000-0001-7145-8674},}
\author[d]{{M.~S.~Wang}\orcidlink{0000-0002-2652-4043},}
\author[e]{{J.~DeRose}\orcidlink{0000-0002-0728-0960},}
\author[f]{{S.~Bailey}\orcidlink{0000-0003-4162-6619},}
\author[g,h,i]{{A.~J.~Ross}\orcidlink{0000-0002-7522-9083},}
\author[f]{{J.~Aguilar},}
\author[j]{{S.~Ahlen}\orcidlink{0000-0001-6098-7247},}
\author[k,l]{{D.~Bianchi}\orcidlink{0000-0001-9712-0006},}
\author[m]{{D.~Brooks},}
\author[n,o]{{F.~J.~Castander}\orcidlink{0000-0001-7316-4573},}
\author[f]{{T.~Claybaugh},}
\author[p]{{K.~S.~Dawson}\orcidlink{0000-0002-0553-3805},}
\author[q]{{A.~de la Macorra}\orcidlink{0000-0002-1769-1640},}
\author[f,r]{{S.~Ferraro}\orcidlink{0000-0003-4992-7854},}
\author[s,t]{{J.~E.~Forero-Romero}\orcidlink{0000-0002-2890-3725},}
\author[n,u,o]{{E.~Gaztañaga}\orcidlink{0000-0001-9632-0815},}
\author[v]{{Satya~{Gontcho A Gontcho}}\orcidlink{0000-0003-3142-233X},}
\author[w]{{G.~Gutierrez},}
\author[x]{{C.~Hahn}\orcidlink{0000-0003-1197-0902},}
\author[y]{{M.~Ishak}\orcidlink{0000-0002-6024-466X},}
\author[z]{{R.~Joyce}\orcidlink{0000-0003-0201-5241},}
\author[z]{{S.~Juneau}\orcidlink{0000-0002-0000-2394},}
\author[f]{{T.~Kisner}\orcidlink{0000-0003-3510-7134},}
\author[f]{{A.~Kremin}\orcidlink{0000-0001-6356-7424},}
\author[i]{{C.~Lamman}\orcidlink{0000-0002-6731-9329},}
\author[f]{{M.~Landriau}\orcidlink{0000-0003-1838-8528},}
\author[f]{{M.~E.~Levi}\orcidlink{0000-0003-1887-1018},}
\author[aa,ab]{{M.~Manera}\orcidlink{0000-0003-4962-8934},}
\author[ac,ab]{{R.~Miquel},}
\author[ad]{{A.~D.~Myers},}
\author[u]{{S.~Nadathur}\orcidlink{0000-0001-9070-3102},}
\author[ae,af,ag]{{W.~J.~Percival}\orcidlink{0000-0002-0644-5727},}
\author[ah]{{I.~P\'erez-R\`afols}\orcidlink{0000-0001-6979-0125},}
\author[ai]{{G.~Rossi},}
\author[aj]{{E.~Sanchez}\orcidlink{0000-0002-9646-8198},}
\author[f]{{D.~Schlegel},}
\author[ak]{{H.~Seo}\orcidlink{0000-0002-6588-3508},}
\author[al]{{G.~Tarl\'{e}}\orcidlink{0000-0003-1704-0781},}
\author[z]{{B.~A.~Weaver},}
\author[f]{{R.~Zhou}\orcidlink{0000-0001-5381-4372},}
\author[am]{{H.~Zou}\orcidlink{0000-0002-6684-3997},}

\affiliation{Affiliations are in Appendix \ref{app:affil}}

\emailAdd{vvaisakh@smu.edu}

\abstract{High-redshift galaxy clustering provides a powerful probe of the growth of structure, testing models of dark matter, dark energy, and galaxy formation during the epoch when the universe was rapidly evolving. Emission line galaxies (ELGs) and quasars (QSOs) are utilized as  tracers of dark matter by the Dark Energy Spectroscopic Instrument (DESI) to probe this redshift regime. We present results from ELG and QSO mock catalogs created from the \Uchuu \textit{N}-body simulation, tuned to DESI Data Release 2 (DR2) \edited{clustering in the range \minFit to \maxFit $\hMpc$}. Employing a modified subhalo abundance matching (SHAM) technique, we populate \Uchuu haloes and subhaloes with QSOs between $0.8 < z < 2.1$. For ELGs, we modify this method to  \edited {only} select satellite galaxies \edited{with low} relative velocities \edited{to their associated central haloes}. We populate a separate set of \Uchuu (sub)halos with ELGs between $0.8 < z < 1.6$. In this paper, we reproduce the redshift evolution of number density and the clustering statistics, across scales ranging from \minScale $\hMpc$ to \maxScale $\hMpc$. We also measure the large-scale clustering bias of both data and mock samples \edited{in the range \minBias to \maxBias $\hMpc$}. These results improve simulated lightcone construction from cosmological models and enhance our understanding of the galaxy-halo connection.

}

\begin{document}
\maketitle
\flushbottom




\begin{center} (DESI Collaboration) \end{center}

\vspace{0.4cm}

\parbox{\textwidth}{
Affiliations are in Appendix~\ref{app:affil}.
}

\date{Accepted XXX. Received YYY; in original form ZZZ}




\section{Motivation}
\label{sec:intro}




The original discovery of dark energy was made through observations of type Ia supernovae (SNe Ia), used as standardizable candles \citep{supernovasearchteam:1998fmf, supernovacosmologyproject:1998vns}, and has since been significantly extended with newer datasets \citep[e.g.][]{JLA, Pantheon, DES3YR, DES5YR, Pantheon+}. Cosmic expansion has also been studied through CMB data, notably from the Planck satellite \citep{Planck2020}. However, a strong discrepancy - currently exceeding a $4\sigma$ significance, has emerged between the local measurement of the Hubble constant ($H_0$) from SNe Ia and the value inferred from the CMB \citep{Verde2019, Freedman:2021ahq, Mortsell:2021nzg}. This Hubble tension continues to prompt investigations into possible systematic effects or indications of new physics \citep[see][and references therein]{Dainotti:2021pqg}. Understanding whether dark energy is a cosmological constant or a manifestation of new physics remains a central challenge in modern cosmology. Alternative methods to measuring cosmic expansion over the widest possible redshift range are critical to understanding the CMB-SNe discrepancy in Hubble Constant measurements, and are also key to commenting on the nature of dark energy.

The large-scale structure (LSS) of the universe offers a powerful probe of cosmic expansion, studied through galaxy clustering measured in large redshift surveys. These structures arise from primordial fluctuations in the early universe, with baryon acoustic oscillations (BAO) imprinting a characteristic scale in the matter distribution. This BAO scale, set at the epoch of recombination, serves as a cosmic standard ruler, allowing precise measurements of the universe’s expansion history. We also use redshift-space distortions (RSD) by quantifying anisotropies in the galaxy power spectrum induced by line-of-sight peculiar velocities. Modeling these anisotropies yields measurements of the growth rate of structure, typically expressed as $f\mathrm\sigma_\mathrm8(z)$, which constrains the expansion history through its dependence on the cosmological model. Combining BAO \& RSD, LSS observations critically complement other cosmological probes such as type Ia supernovae and the cosmic microwave background (CMB), helping constrain cosmological models and the nature of dark energy. Large-scale structure galaxy surveys, such as the Sloan Digital Sky Survey (SDSS; \citep{york2000sdss}), the Dark Energy Spectroscopic Instrument (DESI; \citep{desifdr16a,desifdr16b}), and future facilities including Euclid \citep{laureijs2011euclid} and the Nancy Grace Roman Space Telescope \citep{spergel2015roman}, can provide precise measurements of galaxy clustering and redshift-space distortions.
These are powerful enough to map millions of galaxies over thousands of square degrees, reaching to redshifts well beyond $z=2$. These gains in volume and sampling have driven BAO, RSD, and lensing measurements to unprecedented precision, enabling sharper tests of dark energy and gravity\cite{,DESIDR1BAOConstraints, DESIDR1FullShapeConstraints, DESIDR2Results2, DESFinalCosmologyImplications}.

DESI marks a major advancement in spectroscopic galaxy surveys, delivering unprecedented volumes of data on more than 50 Million galaxies and quasars. Realistic mock catalogs that emulate DESI’s selection functions and statistical characteristics are vital for unlocking its full scientific potential. They enable the testing of cosmological pipelines, assessment of systematic uncertainties, optimization of measurement methods, and validation of theoretical predictions.

\subsection{Modeling the Galaxy-Halo Connection}

Over the past couple of decades, the growing need for realistic and comprehensive mock catalogs has been driven by the advent of major galaxy redshift surveys. These early surveys uncovered the filamentary nature of the universe \citep{colless:2003wz,Alam2021}  - the cosmic web - comprising interconnected galaxies, clusters, and voids. Modeling this complex structure requires accounting for the interplay of gravitational, hydrodynamical, and astrophysical processes, which has presented substantial theoretical and computational challenges. Baryonic physics is substantially more computationally complex than gravitational dark matter dynamics, as it requires modeling nonlinear gas hydrodynamics, radiative cooling, star formation, and feedback processes across a wide range of spatial and temporal scales. As a result, hydrodynamical simulations are limited in volume and resolution, making them impractical for the large survey volumes required by modern cosmological analyses. To meet the precision and volume demands of large-scale structure studies, gravity-only N-body simulations have therefore been developed and widely employed. Notable examples such as the Millennium simulation \citep{springel2005simulations}, the MultiDark simulations \citep{prada2012multidark} and the AbacusSummit simulations \citep{maksimova2021abacussummit} have played a central role in predicting dark matter clustering and providing the foundation for galaxy-halo connection models used in spectroscopic surveys.
When combined with galaxy-halo connection models like the Halo Occupation Distribution (HOD) and Subhalo Abundance Matching (SHAM), these simulations allow for the generation of lightcones that reproduce observed galaxy properties including luminosity, stellar mass, and clustering. 

The Halo Occupation Distribution (HOD) framework statistically models the probability that a dark matter halo of a given mass hosts a specified number of galaxies, separating contributions from central and satellite populations \citep{berlind2002hod,zheng2005hod}. While HOD models are computationally efficient and widely used, Subhalo Abundance Matching (SHAM) directly associates galaxies with resolved dark matter subhalos based on rank-ordered properties, preserving spatial and kinematic information and therefore often providing a more accurate description of small-scale clustering and redshift-space distortions \citep[e.g.][]{marinoni2002mass, kravtsov2004tumultuous, vale2004linking, conroy2006modeling, rodrigueztorres16}. SHAM is a robust approach that links galaxy properties (e.g., stellar mass, luminosity) to dark matter halo properties (e.g., peak circular velocity, $V_\mathrm{max}$) with minimal assumptions about galaxy formation. Unlike HOD models, SHAM assigns galaxies directly to individual halos, naturally incorporating environmental effects. This makes it particularly effective for generating realistic mocks that are well-suited for analyses such as weak gravitational lensing, peculiar velocity studies, and cosmological inference from next-generation surveys. 

In cosmological simulations, lightcones are constructed to mimic the past light path of photons reaching the observer. Unlike snapshots at fixed redshift, lightcones combine outputs from multiple epochs to create continuous, observer-centered catalogs in angular and redshift space. They capture cosmic evolution, encode survey geometry and selection functions, and enable direct comparison with observations. In this work, we leverage simulated lightcones - among the most powerful tools in modern cosmology, particularly for probing the large-scale structure of the universe. These simulations provide crucial insights into the physics governing galaxy and matter evolution, enabling us to test theoretical models, interpret cosmological measurements, and refine observational strategies \citep{ dela-Peacock-2013reconstructing, White2014, rodrigueztorres16, smith17, prada2023desi, dong2024uchuu}.


In this study, our simulation strategy combines large-volume cosmological simulations with flexible empirical models for galaxy - halo connections. We use the Uchuu simulation \citep{ tishiyama2021} for its high resolution and volume, enabling modeling of both small-scale halo physics and large-scale clustering across DESI’s redshift range. We create mock simulations to reproduce the clustering statistics observed in DESI DR1 and DR2 for ELGs and QSOs. Galaxies are assigned to halos using SHAM, which links halo properties to galaxy observables to reproduce number densities and clustering. This approach is designed to produce realistic, high - fidelity mocks that evolve consistently with cosmic time while remaining computationally efficient.  As such, they serve as a critical bridge between theory and observation in the pursuit of a deeper understanding of our universe.

In Section~\ref{sec:desidata} we describe the DESI data samples which we are modeling in these mocks. In Section~\ref{sec:uchuu_simulation} we describe the high-resolution \Uchuu $N$-body simulation that was used to create our simulated lightcones. An overview of the SHAM method adopted to populate ELGs and QSOs into the \Uchuu halo catalogs to build lightcones for each of the DESI tracers is provided in Section~\ref{sec:uchuu_mod_sham}. In Section \ref{sec:paramest} we describe how we fit the modified SHAM mocks to the DESI data. In Section \ref{sec:results} we show results from the best-fit modified SHAM mocks including two point correlation functions, power spectra, halo occupation distributions, and galaxy bias. Finally in Section \ref{sec:concl} we discuss the implications of this analysis including the utility of these mocks in covariance matrices for future DESI analyses. 


\section{DESI \edited{DR1 \& DR2} data}
\label{sec:desidata}
The DESI Survey aims to create the most precise measurements of the cosmic expansion history by using more than 50 million galaxy spectra to map the matter distribution from the local universe up to a redshift of $z  = 3.5$ (\citep{levi2013}). We conduct this survey on the 4m Mayall Telescope at Kitt Peak National Observatory in Arizona. The DESI instrument \citep{desi2022overview} captures spectra from up to 5,000 targets simultaneously using robotic  that direct optical fibers to the target coordinates \citep{silber2023robotic,poppett2024overview, miller2024optical}. Those fibers are then fed into ten three-arm spectrographs covering a wavelength range of 3600-9800 \AA{}.\\
Observations are conducted on sky ``tiles" (~8 square degrees each), with targets assigned to fibers based on priority \citep{schlafly2023survey}. DESI operates distinct observing programs during  ``bright" and ``dark" time, optimizing telescope use by observing brighter objects in its Milky Way Survey (MWS) and Bright Galaxy Survey (BGS) during moonlit conditions and fainter, high-redshift targets (e.g., LRGs, QSOs, ELGs) in darker conditions \citep[]{hahn2023desi,zhou2023target, chaussidon2023target,raichoor2023target}. A ``backup" program runs during poor weather. 
\\
Here we discuss the data taken from the first two DESI main survey data releases. DESI Data Release 1 (\edited{DR1}) includes observations from the first year of DESI’s main survey on the Mayall Telescope at Kitt Peak, Arizona (May 14, 2021 - June 13, 2022) \citep{abdul2025data}. DESI Data Release 2 (\edited{DR2}) uses galaxy catalogs from the first three years of DESI operations (May 14, 2021 - April 2024), which includes DR1 \citep[]{karim2025desi, andrade2025validation}. We present results from both data samples to go with the published \edited{DR1} dataset, and the unpublished \edited{DR2} data to support the \edited{DR2} cosmology analyses using our resulting reference mocks.

During DR1, 2,744 dark-time and 2,775 bright-time tiles were observed. Data were initially processed for quality checks by the DESI spectroscopic pipeline \citep{guy2023spectroscopic} and later refined (``Iron" processing) to produce the DR1 redshift catalogs used in this study.
On the other hand, DESI DR2 contains 6,671 tiles observed during ``dark" time and 5,171 tiles observed during "bright" time, representing 2.4 and 2.3 times the number of tiles released in DR1, respectively.

\subsection{ELG and QSO samples from DR1 and DR2}
\label{sec:ELGQSOSample}
For the observational data, we use the DESI DR2 LSS catalogs \citep{ross2025construction} corresponding to the LOA v2 sample. Table 1 presents the basic properties of the ELG and QSO samples used in this study, including redshift ranges, sky area, total weighted galaxy count, and effective volume. The effective area is computed by generating a large number of random points ($N_\mathrm{random} \sim 10^7$) and determining how many lie within the \edited{DR1} or \edited{DR2} dark footprint for ELGs and QSOs. The effective area, $A$, is then calculated based on this information. $A$ is defined as: 

\begin{equation}
    A = \frac{N_\mathrm{inside}}{ N_\mathrm{random}} A_{\rm full-sky} 
\end{equation}
where $A_{\rm full-sky}$ is the area of the whole sky in square degrees ($4\pi \times \frac{180}{\pi}^2 \sim 41253 \ deg^2$) . $V$ is defined as:
\begin{equation}
    V = \frac{4 \pi}{3} \frac{N_\mathrm{inside}}{ N_\mathrm{random}}  (r_\mathrm{max}^3 -r_\mathrm{min}^3 )
\end{equation}

Here, $r_\mathrm{min}$ and $r_\mathrm{max}$ are the comoving distances corresponding to the redshift limits of each tracer. For the redshift to comving distances transformation, the \Uchuu simulation uses the flat $\LambdaCDM$  with Planck15 parameters: $h=0.6774$, $\Omega_\mathrm{m} = 0.3089$, $\Omega_\mathrm{b} = 0.0486$, $n_\mathrm{s} = 0.9667$, $\Omega_\mathrm{\Lambda} = 0.6911$, and $\sigma_{8} = 0.8159$.

\begin{table}
	\centering
	\begin{tabular}{lcccccc} 
		\hline
		DESI Sample     & Redshift range       & $z_\mathrm{med}$  & $A$ & $N$ & $10^2\times V$  \\
                   &         &       & ($\deg^2$)          &               & ($h^{-3}\mathrm{Gpc}^{3}$)\\
		\hline

        \hline
        \edited{DR1} ELG        & $0.8<z<1.6$ & 1.16  & 5914   & 2432022 & 3.65 \\
        \hline
        \edited{DR2} ELG        & $0.8<z<1.6$ & 1.16  & 10352   & 6534844 & 9.81 \\
        \hline
        \edited{DR1} QSO        & $0.8<z<2.1$ & 1.50  & 7249  & 856652 &  3.14 \\
		\hline
        \edited{DR2} QSO        & $0.8<z<2.1$ & 1.50  & 11181  & 1461588 & 5.36  \\
		\hline
	\end{tabular}
 \caption{Summary statistics for DESI \edited{DR1} and \edited{DR2} samples used - the redshift interval, median redshift ($z_\mathrm{med}$), effective area of the sky footprint weighted by completeness ($A_\mathrm{eff}$), number of galaxies ($N$), and effective volume ($V_\mathrm{eff}$).}
\label{tab:all-basic}
\end{table}

\section{The \Uchuu simulation}
\label{sec:uchuu_simulation}

To model the clustering signal observed in DESI data within a flat $\Lambda$CDM cosmology based on Planck parameters, we employed the \Uchuu $N$-body simulation \citep{tishiyama2021}.
The \Uchuu simulation was executed using the TreePM code \textsc{GreeM} \citep{Ishiyama09, Ishiyama12}. It models a box of comoving length $2~\hGpc$ per side, containing $12,800^3$ dark matter particles. This setup achieves a mass resolution of $3.27 \times 10^8~\hMsun$. Gravitational softening length, a chosen fraction of the inter-particle separation, is set at  $4.27~\hkpc$. Initial conditions were generated using second-order Lagrangian Perturbation Theory (2LPT) at redshift $z_\mathrm{init} = 127$, and the simulation followed the evolution of cosmic structures down to $z = 0$ in the Planck-15 flat $\Lambda$CDM framework.

A total of 50 snapshots spanning $z = 14$ to $z = 0$ were saved. Dark matter haloes and subhaloes were identified using the \textsc{Rockstar} phase-space halo finder \citep{Behroozi13}, and their merger histories were reconstructed using a parallel implementation of the \textsc{ConsistentTrees} algorithm \citep{Behroozi2013b}. Further technical details about the performance of the simulation can be found in Ishiyama et al 2021 \cite{tishiyama2021}. All \Uchuu data products are publicly accessible via the \textsc{Skies $\&$ Universes} platform.\footnote{\url{https://www.skiesanduniverses.org/Simulations/Uchuu/}}. 

For each halo and subhalo, we determined the peak circular velocity, \vpeak - defined as the highest value of $V_\mathrm{max} = \rm{max}(\sqrt{GM(r)/r})$ measured across all 50 redshift outputs. This \vpeak metric forms the basis for the SHAM implementation, serving as a proxy for galaxy luminosity or stellar mass in assigning galaxies and quasars to simulated halos. Of the three dark time tracers in DESI (LRGs, ELGs, and QSOs), LRGs are the only ones which are believed to be complete in halo mass occupation \citep{Alam_2020}, so they are the only tracers which can use a traditional abundance matching technique. ELGs and QSOs are not complete in any known parameter space (\citep{Favole16,rodrigueztorres17}), so the traditional SHAM method must be modified. We describe the modified SHAMtechniques used in this analysis in Section \ref{sec:uchuu_mod_sham}.


In this work, we adopt the peak maximum circular velocity \vpeak as a proxy for (sub)halo mass. \vpeak has been extensively used as a halo mass proxy in numerous abundance matching studies to accurately reproduce the properties of observed galaxies in large-scale surveys \citep[]{Conroy06, trujillo-gomez11, Nuza13, Reddick13, chaves-montero16, rodrigueztorres16, safonova21}.

The resulting Uchuu halo catalogs enable us to populate mock galaxy catalogs. We do this using a Subhalo Abundance Matching (SHAM) technique and DESI Survey data. The resulting mock lightcones effectively reproduce the observed number densities and clustering patterns of each DESI tracer. A detailed account of the lightcone construction process is provided in our paper using the One percent Survey data \citep{prada2023desi}. In Section~\ref{sec:results}, we compare the clustering predictions from the mocks with the measurements from the DESI DR1 and DR2, allowing us to study the halo occupation distribution and large-scale bias of ELGs and QSOs. We applied the aforementioned steps to the ELG and QSO tracers to create their respective full-sky lightcones. The lightcones were then cut to match the northern and southern areas of the DESI data. 

\begin{table*}\hspace*{-1.cm}
\centering
\begin{tabular}{cccccccccc}
    \hline
    \vspace{2pt}
    $z$ range & $N$ & $V$ & $n_{\rm g}^{\rm ELG}$   &  $V_\mathrm{mean}$ & $f_{\rm sat}$ & $b^{\rm ELG}$ & $b^{\rm Uchuu}$ \\
     &  & $\edited{Gpc^{3}\cdot h^{-3}}$ & $\edited{Gpc^{-3}\cdot h^{3}}$  &  $\edited{km \cdot s^{-1}}$ & &  & \\
    \hline
    0.8--1.6 & 15094649 & 20.28  & $\edited{5.5\times10^{-4}}$ &  206.6$\pm$1.5 & 0.318$\pm$0.007 & 1.325$\pm$0.002 & 1.296$\pm$0.002 \\
    0.8--1.1 &  6271380 & 6.41   & $\edited{7.2\times10^{-4}}$ & 193.8$\pm$2.4 & 0.357$\pm$0.090 & 1.177$\pm$0.003 & 1.126$\pm$0.003 \\
    1.1--1.3 & 4088265  & 5.18   & $\edited{5.9\times10^{-4}}$ &   207.5$\pm$2.7 & 0.309$\pm$0.080 & 1.312$\pm$0.004 & 1.302$\pm$0.004 \\
    1.3--1.6 & 4735004  & 8.69   & $\edited{4.1\times10^{-4}}$ &  220.7$\pm$2.7 & 0.288$\pm$0.090 & 1.494$\pm$0.003 & 1.500$\pm$0.002 \\
    \hline
\end{tabular}

\caption{Best-fit modified SHAM parameters for the ELGs for each redshift bin and for the entire mock. Columns provided in order: Redshift range of different ELG samples, number of galaxies, effective volume (in $h^{-3}\mathrm{Gpc}^{3}$) and galaxy number density (in $h^{3}\mathrm{Mpc}^{-3}$).
The last two columns show the bias calculated from the data and the mocks respectively ($b^{\rm ELG}$ and $b^{\rm Uchuu}$). The errors reported for the data are statistical.} 
\label{tab:elg-shamparm}
\end{table*}


\begin{table*}\hspace*{-1.cm}
\centering
\begin{tabular}{cccccccccc}
    \hline
    \vspace{2pt}
    $z$ range & $N$ & $V$ & $n_{\rm g}^{\rm QSO}$ &  $V_\mathrm{mean}$ & $f_{\rm sat}$ & $b^{\rm QSO}$ & $b^{\rm Uchuu}$ \\
    &  & $\edited{Gpc^{3}\cdot h^{-3}}$ & $\edited{Gpc^{-3}\cdot h^{3}}$  &  $\edited{km \cdot s^{-1}}$ & &  & \\
    \hline
    0.8--2.1 & 1944628 & 36.15  & $3.55 \times 10^{-5}$ &  313$\pm$14 & 0.137$\pm$0.021 & 2.215$\pm$0.003 & 2.175$\pm$0.006 \\
    0.8--1.1 & 321006  & 6.38   & $3.32 \times 10^{-5}$ &  309$\pm$10 & 0.095$\pm$0.053 & 1.683$\pm$0.011 & 1.638$\pm$0.011 \\
    1.1--1.4 & 465794  & 7.97   & $3.85 \times 10^{-5}$ &  311$\pm$16 & 0.128$\pm$0.047 & 1.915$\pm$0.008 & 1.932$\pm$0.006 \\
    1.4--1.7 & 530389  & 8.98   & $3.90 \times 10^{-5}$ &  313$\pm$14 & 0.140$\pm$0.046 & 2.363$\pm$0.006 & 2.292$\pm$0.008 \\
    1.7--2.1 & 627439  & 12.81  & $3.23 \times 10^{-5}$ &  315$\pm$18 & 0.208$\pm$0.054 & 2.763$\pm$0.005 & 2.699$\pm$0.010 \\
    
    \hline
\end{tabular}
\caption{Same as Table \ref{tab:elg-shamparm} but for QSOs. The last two columns show the bias calculated from the data and the mocks respectively ($b^{\rm QSO}$ and $b^{\rm Uchuu}$).  }
\label{tab:qso-shamparm}
\end{table*}

\begin{figure}
	\includegraphics[width=\columnwidth]{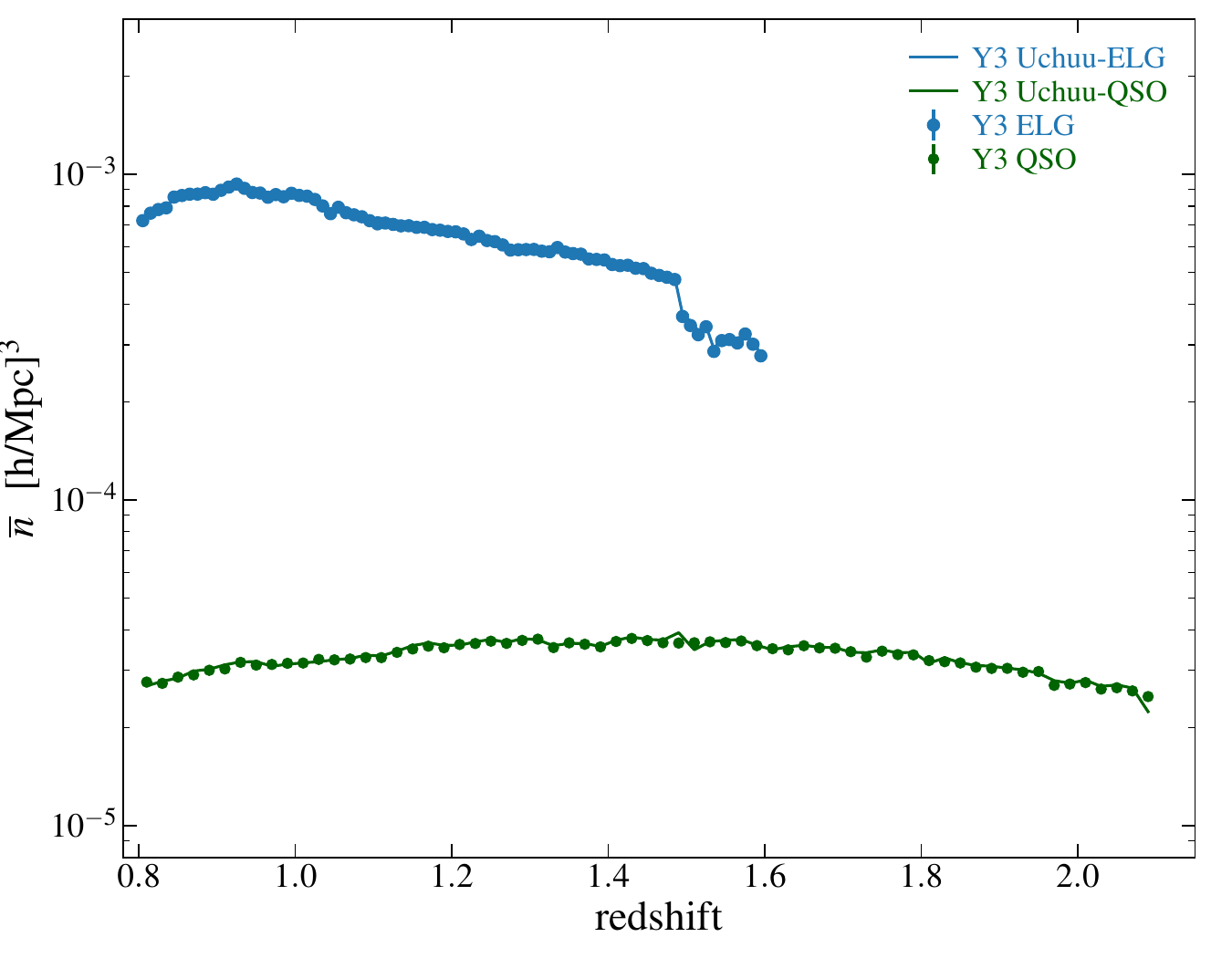}
    \caption{The comoving number density of the ELG and QSO samples (points) and the corresponding \Uchuunospace-DESI mock lighcones (solid line) over the entire redshift range $0.8 < z < 2.1$. 
    }
    \label{fig:all-ndens}
\end{figure}

Figure \ref{fig:all-ndens} shows a comparison between the comoving number density of the DESI data (points) and the mean comoving number density of the \Uchuu DR2 mock lightcones (solid lines). We show agreement with the data n(z) as expected since we use the data n(z) to generate the mocks.

\section{Modified (Sub)halo Abundance Matching (SHAM)}
\label{sec:uchuu_mod_sham}

We employ a modified Subhalo Abundance Matching (SHAM) framework to model the halo occupation and clustering of DESI ELGs and QSOs. Our approach \edited{is based on the method} previously applied to the BigMultiDark Planck simulations \citep{klypin2016multidark} to construct eBOSS \citep{dawson2016eboss} QSO lightcones (see Sections 3.1 and 3.2 of \citep{rodrigueztorres17}). \edited{This method} adapts traditional abundance matching techniques for use with incomplete tracers like ELGs and QSOs. A similar strategy has also been used for modeling [OII] emitters \citep{Favole2017}, where incompleteness of tracer populations in baryonic properties such as luminosity and stellar mass is explicitly accounted for.

\edited{Our implementation of this model begins by constructing Gaussians to serve as the final \vpeak distributions for the (sub)haloes that will host ELGs or QSOs. These Gaussian distributions are defined by two free parameters each: their means ($V_{\rm mean, c/s}$) and their standard deviations ($\sigma_{\rm V, c/s}$) - where \textit{c} refers to Central galaxies, and \textit{s} refers to Satellite galaxies respectively.  The fraction of each tracer type that are located in satellites ($f_{\rm sat}$) serves as an additional free parameter that controls the relative normalization of the Gaussians for the centrals and satellites. To further simplify the model, we fix the means and standard deviations of the central and satellite galaxy \vpeak distributions to be the same values  ($V_{\rm mean}$ and $\sigma_{\rm V}$), reducing the number of free parameters from 5 to 3. This makes the final \vpeak distribution of ELGs or QSOs:} 

\begin{equation}
\label{eq:modsham1}
\begin{split}
     \phi_{\rm ELG/QSO}(V_{\rm peak}) 
     = & \hspace{3pt} \mathcal{G}_{\rm s}(V_{\rm peak}; V_{\rm mean}, \sigma_V)  \\
    &  + \mathcal{G}_{\rm c}(V_{\rm peak}; V_{\rm mean}, \sigma_V).
\end{split}
\end{equation}

\edited{We normalize the Gaussian distributions to the number density observed in each redshift bin (dz = 0.02) of the DESI data ($\rho(z)$) and the comoving volume ($V_{\rm C}(z)$) from the Uchuu fiducial cosmology as shown in equation \ref{eq:modsham2} below: }

\edited{
\begin{equation}
\label{eq:modsham2}
\begin{split}
     \int_0^\infty  \mathcal{G}_{\rm s}(V_{\rm peak}, z_{\rm cent}; V_{\rm mean}, \sigma_V) dV_{\rm peak}=&  \hspace{3pt} (V_{\rm c}(z_{\rm max}) - V_{\rm c}(z_{\rm min}))*\rho(z_{\rm cent})*f_{\rm sat} \\ 
    \int_0^\infty  \mathcal{G}_{\rm c}(V_{\rm peak}, z_{\rm cent}; V_{\rm mean}, \sigma_V)dV_{\rm peak} =&  \hspace{3pt} (V_{\rm c}(z_{\rm max}) - V_{\rm c}(z_{\rm min}))*\rho(z_{\rm cent})*( 1 - f_{\rm sat})
\end{split}
\end{equation}
}
\noindent
\edited{where $z_{\rm min}$ and $z_{\rm max}$ are the edges of the redshift bin centered at $z_{\rm cent}$.}

\edited{Finally, we use these Gaussians to determine the probability for each (sub)halo \edited{within a lightcone of Uchuu halos (see \S 4.3 for Lightcone construction method)}  to be selected to host an ELG or QSO. This is done within redshift bins \edited{(with edges $z_{\rm min}$ and $z_{\rm max}$)} and \vpeak bins \edited{(with edges $V_{\rm lower}$ and $V_{\rm upper}$)} as shown in equation \ref{eq:modsham3} below: }
\edited{
\begin{equation}
\label{eq:modsham3}
    P_{\rm c/s}(V_{\rm lower} <V_{\rm peak} < V_{\rm upper}; z_{\rm min} < z < z_{\rm max}) = \frac{\int_{V_\mathrm{lower}}^{V_\mathrm{upper}} \mathcal{G}_{\rm c/s}(V_{\rm peak}, z) dV_{\rm peak}}{\Sigma_{z_{\rm min}}^{z_{\rm max}}\Sigma_{V_{\rm min}}^{V_{\rm max}}N_{\rm c/s}}
\end{equation}
}
\noindent
This method, \edited{using a Monte Carlo rejection sampling,} enables a more flexible and accurate reproduction of the observed ELG and QSO populations, especially where the traditional SHAM framework is limited by assumptions of completeness.

\subsection{QSO Lightcone}
\label{sec:uchuu_qso}

\noindent We generate a grid of full sky QSO lightcone mocks in this parameter space, compute the monopole of the two-point correlation function (2PCF) for each of these mocks, and compute a $\chi^2$ statistic for each 2PCF monopole with respect to that of the DESI data, in the separation range  $\sim 3~\hMpc$ to $\sim 70~\hMpc$, using the square root of the diagonal of the ($N=128$) jackknife covariance matrix of the data 2PCF as the uncertainty. \edited{The grid of parameters spans a range of $V_{\rm mean}$ of 270 - 390 km/s sampled every 15 km/s, a range of $f_{\rm sat}$ from 0\% to 40\% sampled every 2.5\%, and a single $\sigma_{V}$ value of 30 km/s. We only sample a single $\sigma_{V}$ value as in \citep{rodrigueztorres17} due to the relatively small impact of $\sigma_{V}$ on the mock clustering within the fit range relative to the other parameters as shown in Figure \ref{fig:sigmavary}.} The best fit mock parameters were determined by first finding the minimum $\chi^2$ value over the grid of parameters and then fitting a 2 dimensional paraboloid to the $\chi^2$ values vs. \edited{$V_{\rm mean}$ and $f_{\rm sat}$}.
The best fit $V_\mathrm{mean}$ and $f_\mathrm{sat}$ parameters
are listed in Table~\ref{tab:qso-shamparm}, used to generate our \Uchuu lightcones for QSO.   We work with the (sub)halo catalogues from \Uchuu boxes at five different redshifts, namely $z =$ 0.86, 1.03, 1.32, 1.65, and 1.9  - and create shells to cover the redshift range of the QSO sample (0.8 to 2.1). Estimates of quasar redshift have large uncertainties \citep{Chaussidon23} of a few hundred $\kms$ due to the broadness of the emission lines and the intrinsic shifts from other emission lines  \citep{youles2022effect}. Hence we introduce Gaussian redshift errors such that
\begin{equation}
    z_\mathrm{final} = z + \mathcal{G}(0, \sigma).
\end{equation}
Here, $z_\mathrm{final}$ is the final redshift distribution for the mock quasar catalogs, and $\mathcal{G}(0, \sigma)$ is the Gaussian random error added to the initial redshift distribution $z$. Although the dispersion $\sigma$ was set as a constant $500~km/s$ for the One-Percent sample over all redshifts, for DR1 and DR2, we follow the results from Figure 8 in the DESI Mg II Absorber analysis \citep{napolitano2023detecting}.
\\

\subsection{ELG Lightcone}
\label{sec:uchuu_elg}

\noindent Emission line galaxies present a unique opportunity in our data sample because their high number density permits an examination of the clustering at small scales.  Conversely, the physics of star formation means that we may have a strongly biased sample of galaxies making into the ELG target selection used in the DESI survey.  In particular, we have reason to anticipate that those satellites that are fast moving in their dark matter haloes will tend strongly to not exhibit strong star formation when they are close to their central galaxies. The ram pressure from the denser local environment will strip out gas in the satellite and quench star formation \citep{boselli2022ram}.  We will not observe these galaxies as ELGs and they should be omitted from the sample.  We therefore modify our ELG modeling to better account for this by introducing two new parameters.  $V_\mathrm{max}/V_\mathrm{peak}$ is designed to require satellites have maintained their velocity over their development.  More importantly, we consider that both the  central peak circular velocity, and the relative velocity of the satellite with respect to this central should be low, to reduce stripping. This will be explored further in future analyses \citep{Amalbert2026UchuuELG}.  We implement this selection as
\begin{itemize}
    \item[(i)] $V_\mathrm{max} / V_\mathrm{peak} > 0.9, $ (to avoid major mergers).
    \item[(ii)] $V_\mathrm{peak}^\mathrm{Central}  < 300  km/s, \And  (V_\mathrm{xyz,pec.} - V^\mathrm{Central}_\mathrm{xyz,pec.}) < 300 km/s $ (Slow-moving Satellite selection).
\end{itemize}

From \Uchuu, we select halos based on the above thresholds. Then we follow the same procedure as the QSO lightcone selection described in \ref{sec:uchuu_qso}. 
\edited{The grid of parameters spans a range of $V_{\rm mean}$ of 180 - 250 km/s sampled every 5 km/s, a range of $f_{\rm sat}$ from 0\% to 45\% sampled every 2.5\%. As with the QSO light cone fitting, we only sample a single $\sigma_{V}$ value of 30 km/s.} The best-fit $V_\mathrm{mean}$ and $f_\mathrm{sat}$ parameters used to generate our ELG mock are listed in the first row of Table~\ref{tab:elg-shamparm}. The subsequent rows show the best fit parameters for each box used in the construction of the mocks fit separately.  We apply the modified SHAM method to the (sub)halo catalogues from \Uchuu boxes at redshifts 0.94, 1.22 and 1.43 to cover the interval $0.8 < z < 1.6$.  
\\

\subsection{Construction of Lightcones}

After starting with the relevant snapshots from \Uchuu boxes , we generate the Uchuu-DESI lightcones for ELGs and QSOs using the method described in Smith et al \citep{smith22b}. This involves replicating the cubic Uchuu boxes to cover the required volume and subsequently cutting them in redshift space to produce spherical shells. To build lightcones, we center boxes on the observer and replicate them in a volume that allows extraction of a spherical shell of a given redshift and thickness.  Cartesian (x,y,z) coordinates are converted to (RA, DEC, Redshift $z$) incorporating the effects of peculiar velocities, including line-of-sight effects like Fingers-of-God.  The replications are periodic and created to guarantee we fill the spherical shell centered on specific Uchuu snapshot redshifts. These shells are then merged to construct the full lightcone corresponding to the DESI survey volume.  Building the lightcone is done in a way that ensures the correct number density of galaxies/quasars and prevents discontinuities at the interfaces between shells. 

To match the observed ${\Delta}z$ in each redshift bin, we select only the required number of galaxies to ensure the resulting $n(z)$ aligns with the observed distribution in all redshift bins. The bin width is set to ${\Delta}z= 0.01$ for both ELGs and QSOs.

\begin{figure}
	\includegraphics[width=\columnwidth]
    {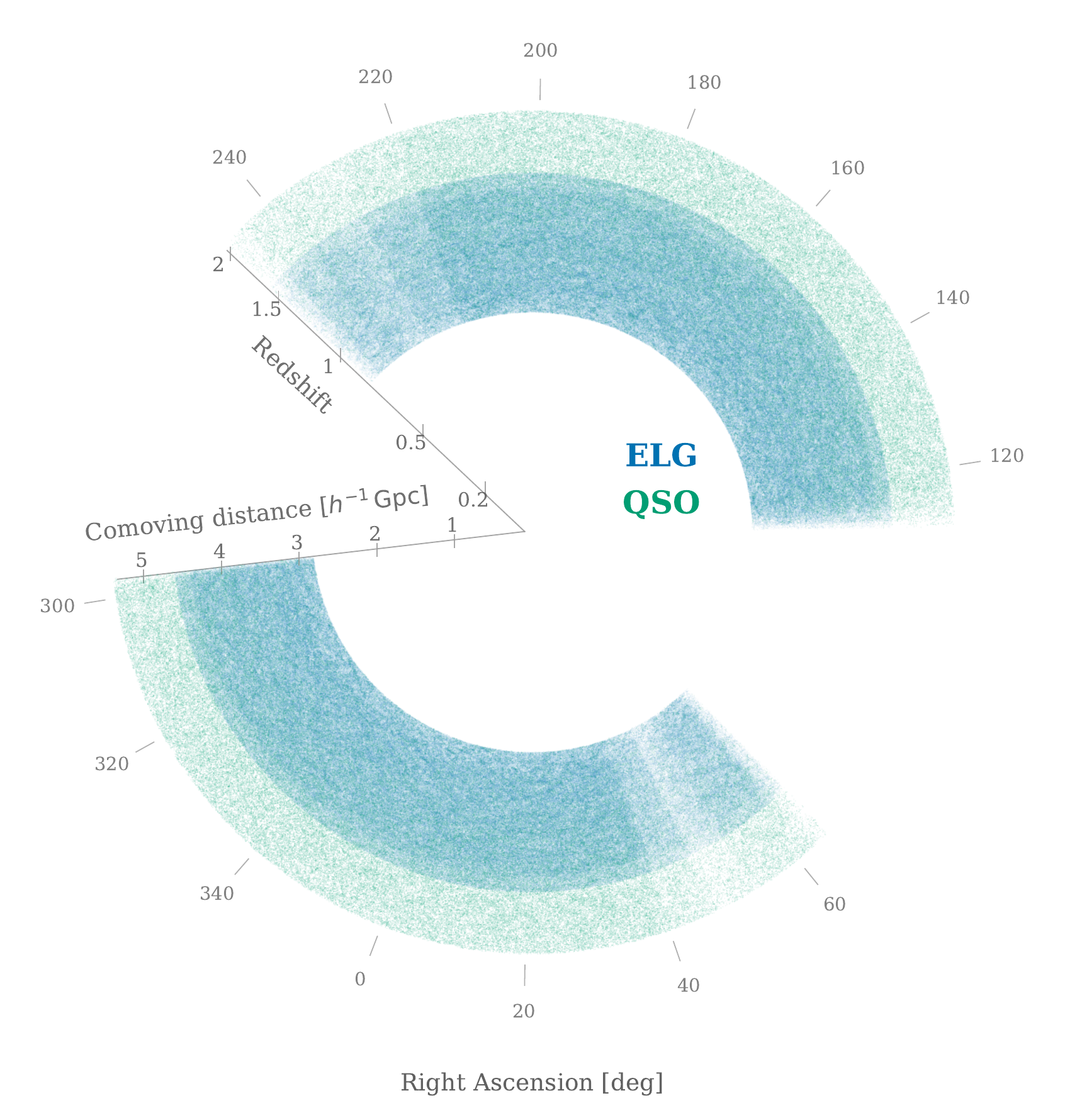}
    \caption{A slice of the universe mapped by Uchuu DR2 drawn from a small wedge of the DESI footprint between $±5$ degrees in
declination out to $z = 2$. We render emission line galaxies (ELG), and QSOs using blue and green points, respectively. The shade of the colour maps to declination (lighter colours correspond to higher declination). 
    }
    \label{fig:uchuu-desi}
\end{figure}

ELG lightcones spanning $0.8 < z < 1.6$ were produced using the following Uchuu boxes: $z$ = 0.94 ($0.8 < z < 1.1$), 1.22 ($1.1 < z < 1.3$), and 1.43 ($1.3 < z < 1.6$). QSO lightcones spanning $0.8 < z < 2.1$ were produced using the following Uchuu boxes: $z$ = 0.86 ($0.8 < z < 0.9$), 1.03 ($0.9 < z < 1.2$), 1.32 ($1.2 < z < 1.5$), 1.65 ($1.5 < z < 1.8$) and 1.90 ($1.8 < z < 2.1$). Although not our primary goal, we also attempt to construct a higher redshift extension to this lightcone, using $z$ = 2.3 ($2.1 < z < 2.8$) and 3.31 ($2.8 < z < 3.5$) - by extending the best-fit model. 
For both tracers, We follow it up by applying the DESI DR2 footprint mask on the full catalog.

Figure~\ref{fig:uchuu-desi} presents a visual representation of all four DESI tracers (including the BGS and LRG \citep{fernandez2025desi}) within a slice of an \Uchuunospace-DESI lightcone.

\section{Best-fit parameter estimation}
\label{sec:paramest}
Once we obtain the lightcones, we then compute the 2PCF and compare them with the DESI data. We construct a grid of full-sky ELG and QSO lightcone mock catalogs across a defined parameter space. For each mock, we compute the monopole of the two-point correlation function (2PCF). To evaluate how well each mock reproduces the clustering observed in the DESI data, we calculate a $\chi^2$ statistic by comparing the mock 2PCF monopole to that of the data over the separation range of $3h^{-1}\text{Mpc}$ to $70h^{-1}\text{Mpc}$. The uncertainties in this comparison are estimated using the square root of the diagonal elements of the \edited{($N=128$)} jackknife covariance matrix derived from the data.

To determine the best-fit model parameters, we we first examine the variation in $\chi^2$ for each parameter ($V_\mathrm{mean}$, $\sigma_{v}$, $f_\mathrm{sat}$)) keeping the other two fixed.  
We then construct the 2-D distribution of $V_\mathrm{mean}$ vs. $f_\mathrm{sat}$ and fit the $\chi^2$ to a paraboloid \edited{using \textsc{MINUIT} \cite{MINUIT,IMINUIT}} to obtain the optimal point \edited{after excluding values more than 30 from the lowest $\chi^2$ value where the parabolic approximation no longer applies}.  We then refine this estimate by fitting a parabola to the one-dimensional slice of this $\chi^2$ distribution.  We do this independently for each parameter ($V_\mathrm{mean}$, $f_\mathrm{sat}$).
The optimal values of $V_\mathrm{mean}$ and $f_\mathrm{sat}$ are taken as the vertices of these fitted parabolas \edited{ and their uncertainties are taken as the change in each value from the minimum to increase the $\chi^2$ by 1 as calculated by \textsc{MINUIT}}. An example is shown in the Appendix~\ref{app:heatmaps-best-fit}.

\begin{figure*}
    \includegraphics[width=0.49\columnwidth]{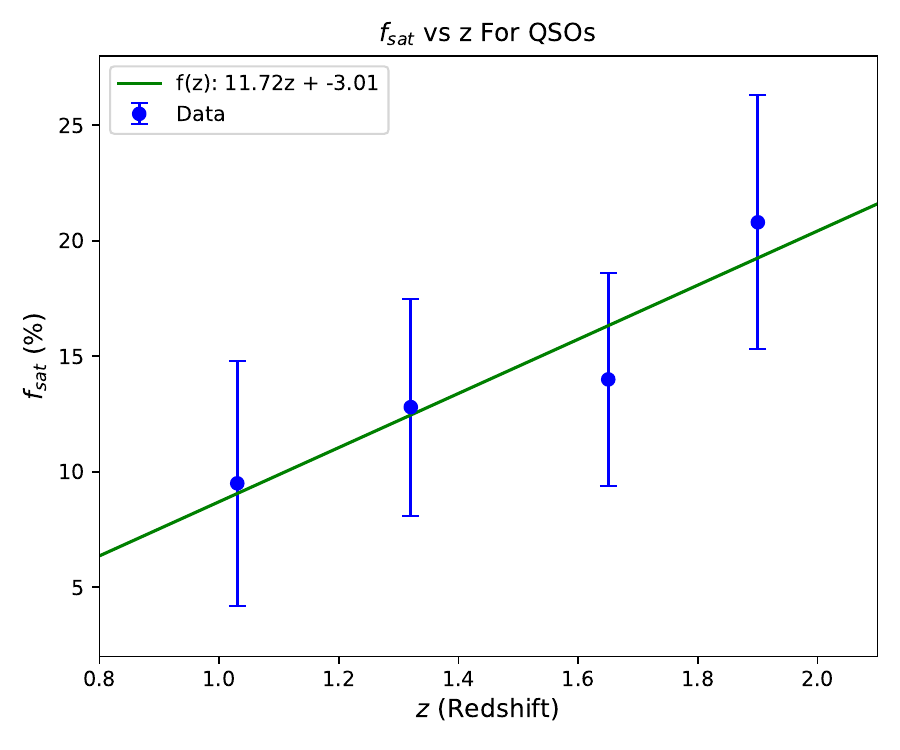}
    \includegraphics[width=0.49\columnwidth]{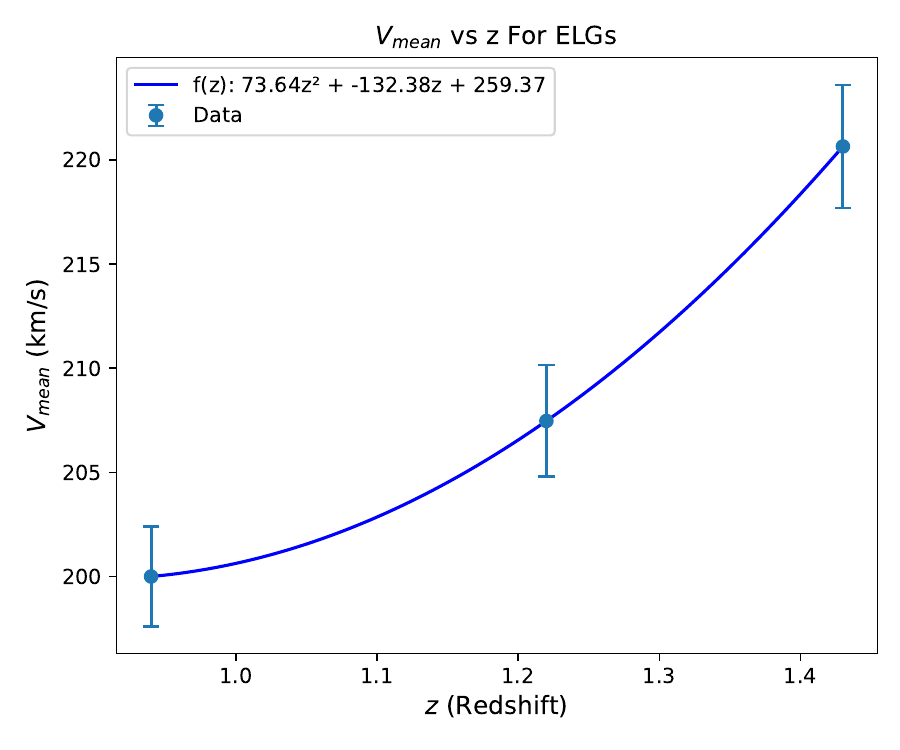}
    \hspace{0.02\textwidth}
    
    \caption{Modelling the QSOs (left panel) and the ELGs (right panel) as a function of redshift. The QSOs seem to have a linearly increasing trend in satellite fraction $f_\mathrm{sat}$ with redshift, and a fixed $V_\mathrm{mean}$ of 313 $km/s$. We also predict the mock behaviour at higher redshifts using this trend. Although it was not included in the fitting range, we also produce a higher redshift extension to the QSO mock in the $ 2.1 < z < 3.5$ regime.  This is described by a function: $f(z) = 11.72z - 3.01$ . On the other hand, the ELGs \edited{do not show} variation with $f_\mathrm{sat}$ (fixed at 31 \%), but \edited{do show} an increasing $V_\mathrm{mean}$ with redshift, described by: $f(z) = 73.64z^{2} - 132.38z + 259.37$.  \edited{We chose to proceed with a quadratic interpolation rather than a linear fit for the ELG $V_{\rm mean}$ values because the parameters obtained from fitting to a straight line produced lower clustering for the first redshift bin.}} 
    \label{fig:elg_vmean_z}
\end{figure*}

From the best-fit values obtained in each redshift bin (see Tables~\ref{tab:elg-shamparm} and~\ref{tab:qso-shamparm}), we identify a clear redshift dependence in the SHAM parameters. In particular, the satellite fraction $f_{\mathrm{sat}}$ exhibits redshift evolution for QSOs, while the mean velocity parameter $V_{\mathrm{mean}}$ shows redshift dependence for ELGs, as illustrated in Figure~\ref{fig:elg_vmean_z}. This behavior enables a stable and physically motivated parameterization of the model, in which the redshift dependence is captured explicitly rather than treated independently in each bin. Consequently, instead of fitting three independent parameters in each of four redshift bins (a total of 12 free parameters), the QSO model can be effectively described by a reduced parameter set: a redshift-dependent $f_{\mathrm{sat}}(z)$, along with redshift-independent $V_{\mathrm{mean}}$ and $\sigma_v$. This reduction represents a substantial improvement over earlier implementations of this method, both in efficiency and interpretability.

\section{Results}
\label{sec:results}

In this section, we compare the DESI ELG and QSO clustering signal with that predicted by the Planck cosmology using our \Uchuu DR1 and DR2 mock lightcones, as described in the previous section. We explore the dependence of the galaxy clustering on redshift for ELGs and QSOs,  and estimate the halo occupancy and large-scale bias for both targets.

\subsection{Clustering statistics: DESI vs. \Uchuu}
\label{sec:results-clustering}

\subsubsection{Redshift Space Two-Point Correlation Function}

We use the Landy-Szalay \citep{LS93} estimator to measure the two-dimensional correlation function, $\xi(s,\mu)$,in bins of $s$ and $\mu$, 
\begin{equation}
    \xi(s,\mu) = \frac{DD(s,\mu) - 2DR(s,\mu) + RR(s,\mu)}{RR(s,\mu)}
\end{equation}
\noindent where the normalized pair counts in the correlation function estimate consist of $DD$, representing the counts of data galaxies with other data galaxies, $DR$ indicates counts of data galaxies with random points, and $RR$ are the counts of random points with other random points. $s$ represents the separation between a pair of objects in units of $\hMpc$, and $\mu$ is the cosine of the angle between the pair separation vector and the line-of-sight. We used logarithmically spaced bins of separation and linearly spaced bins in $\mu$ between -1 and 1. We used 49 separation bins between 0.01 and 100 $\hMpc$ and 201 $\mu$ bins. The random catalogs for the mock samples were generated by uniformly populating the survey footprint and redshift range, following the same angular and radial selection functions as the data.

We decompose $\xi(s,\mu)$ into Legendre polynomials,
\begin{equation}
\xi_\ell(s) = \frac{2\ell+1}{2} \int^1_{-1} \xi(s,\mu)P_\ell(\mu)d\mu .
\end{equation}
We measure the monopole and quadrupole ($\ell=0,2$), which are the first non-zero Legendre multipoles of the redshift-space two-point correlation function. To account for the selection function, we generate uniform random samples that are 20 times larger than the survey data and use them to estimate the data-random and random-random pair counts for each tracer. We estimate the two-point correlation functions using the Python package \textsc{pycorr}\footnote{https://github.com/cosmodesi/pycorr} (a modified version of \textsc{Corrfunc} \citep{Corrfunc2020}).


The data sample is corrected for incompleteness (e.g. due to fiber collision effects and a finite number of tilings) using the Pairwise Inverse Probability (PIP) weights developed by Bianchi and Percival 2017 \cite{BianchiAndPercival} combined with the angular upweighting method developed by Percival and Bianchi 2017 \cite{PercivalAndBianchi}. Both of these weights are determined using a set of 128 alternate realizations of DESI fiber assignment known as Alternate Merged Target Ledgers (AMTLs)\cite{lasker24a}. The combination of PIP weighting and angular upweighting was chosen since it has been shown to produce a practically unbiased estimator of the true two point clustering statistics even when pairs of galaxies in the parent catalog are never observed in any of the AMTL realizations (i.e. zero probability pairs). 

Additionally, the data and mock TPCFs are weighted by FKP weights \cite{fkp} to account for variable galaxy number density with redshift and to minimize variance. FKP weights are defined as
\begin{equation}
w_\mathrm{FKP} = \frac{1}{1+P_0 n(z)},
\end{equation}
where $n(z)$ is the weighted number density, and $P_0$ is the power spectrum value at a scale of $k = 0.15~h\mathrm{Mpc}^{-1}$ where we desire to minimize the variance. We adopt $P_0 = 4000~h^{-3}\mathrm{Mpc}^3$ for ELG, and $P_0 = 6000~h^{-3}\mathrm{Mpc}^3$ for QSO \citep{dr}.



\begin{figure*}

    \includegraphics[width=0.52\columnwidth]{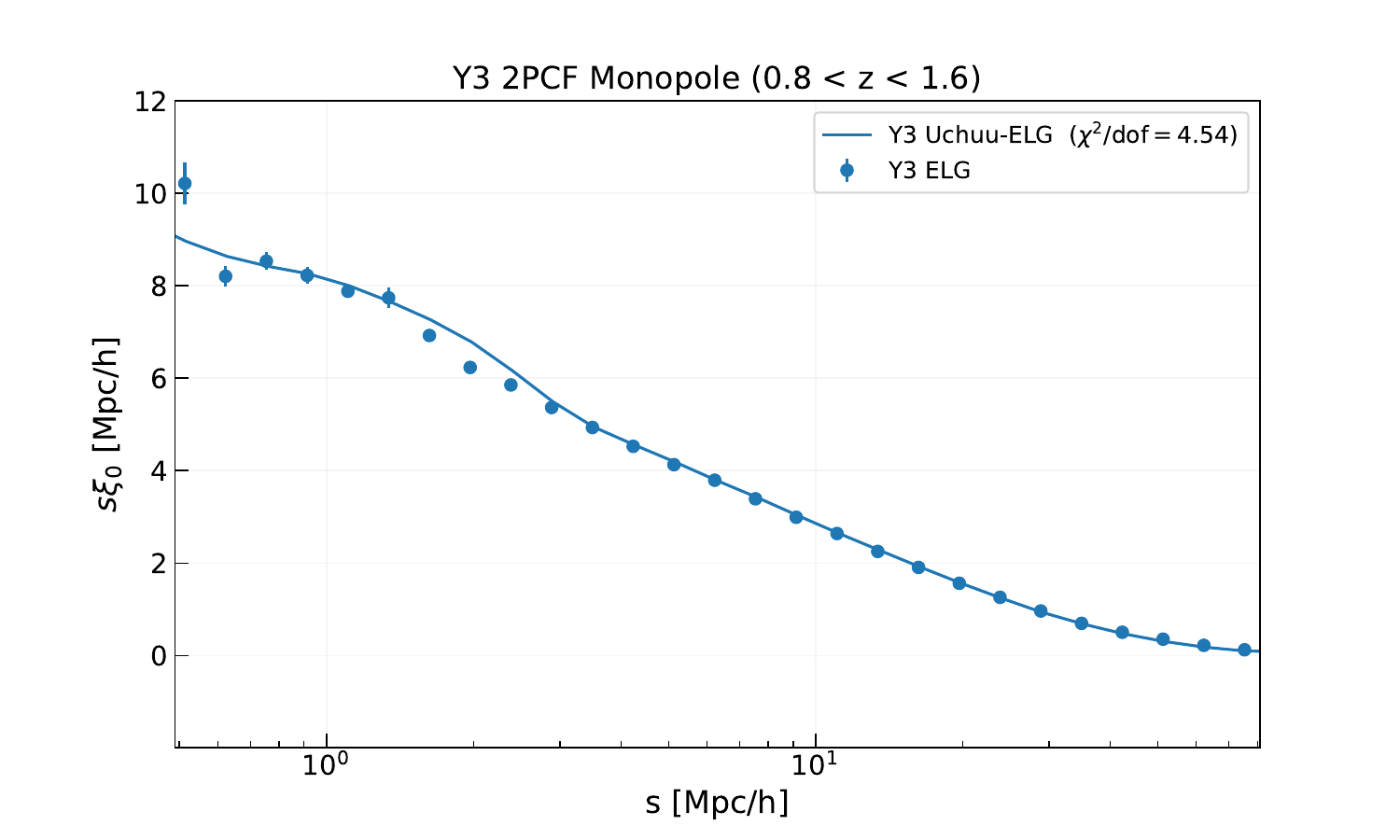}
    \includegraphics[width=0.52\columnwidth]{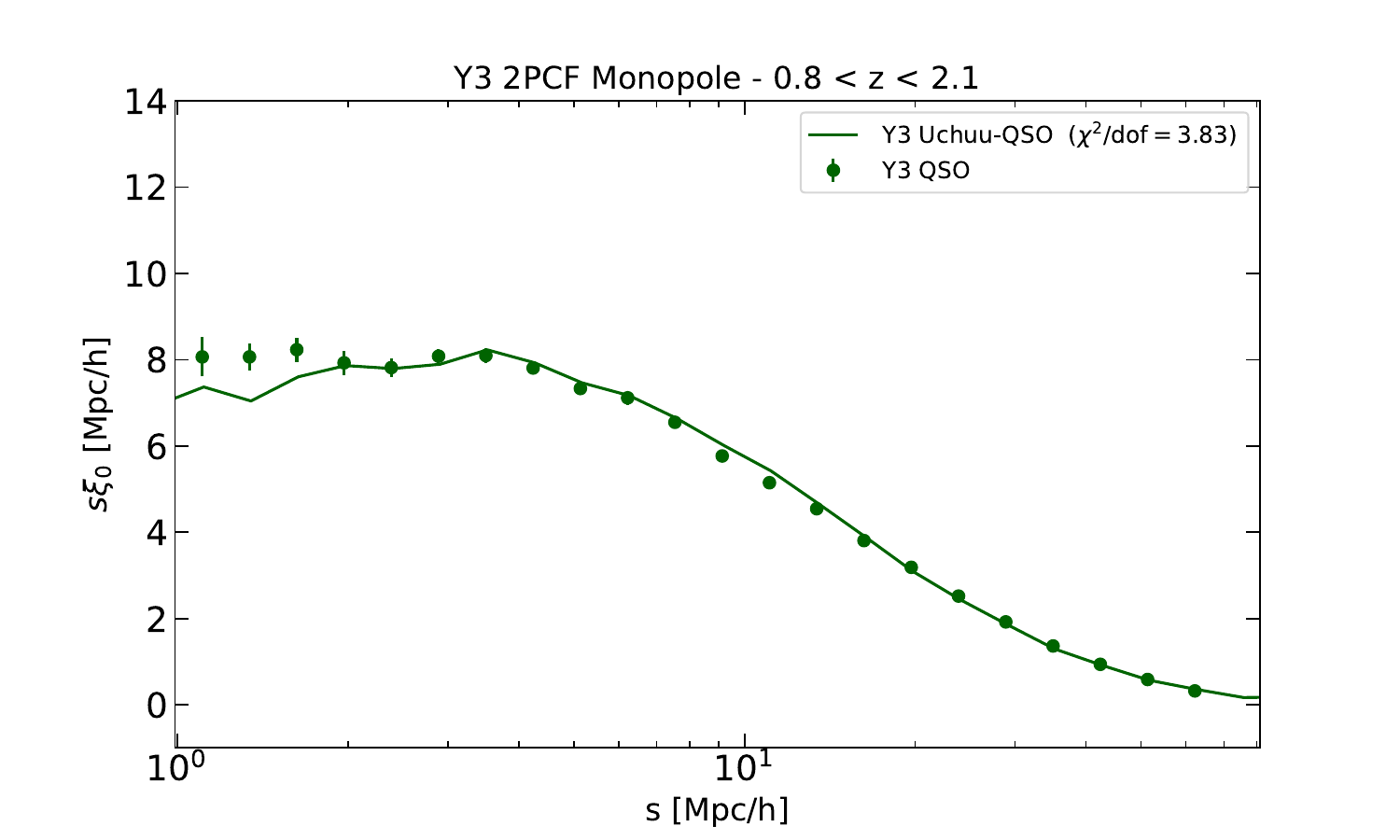}
    
    \caption{Measurements of the monopole of the redshift-space correlation function for $0.8 < z < 1.6$ (ELGs) and $0.8 < z < 2.1$ (QSOs).  The points with error bars represent the measurements from the DESI DR2.
    The results show agreement between theory and observation within the fit range for the modified SHAM models (from $3 \hMpc$ to $70 \hMpc$).}
    \label{fig:sv3-2pcf}
\end{figure*}

\begin{figure*}

    \includegraphics[width=0.52\columnwidth]{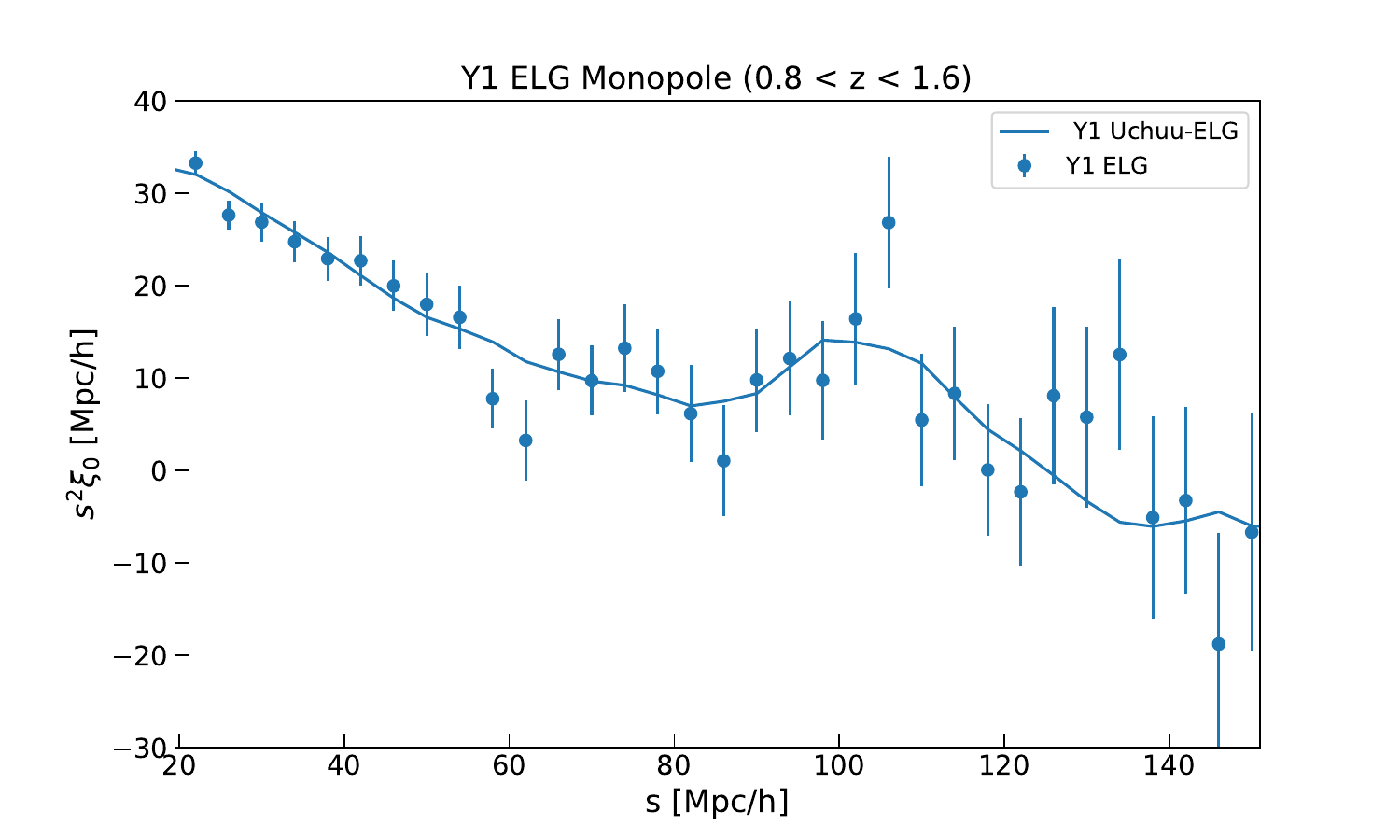}
    \includegraphics[width=0.52\columnwidth]{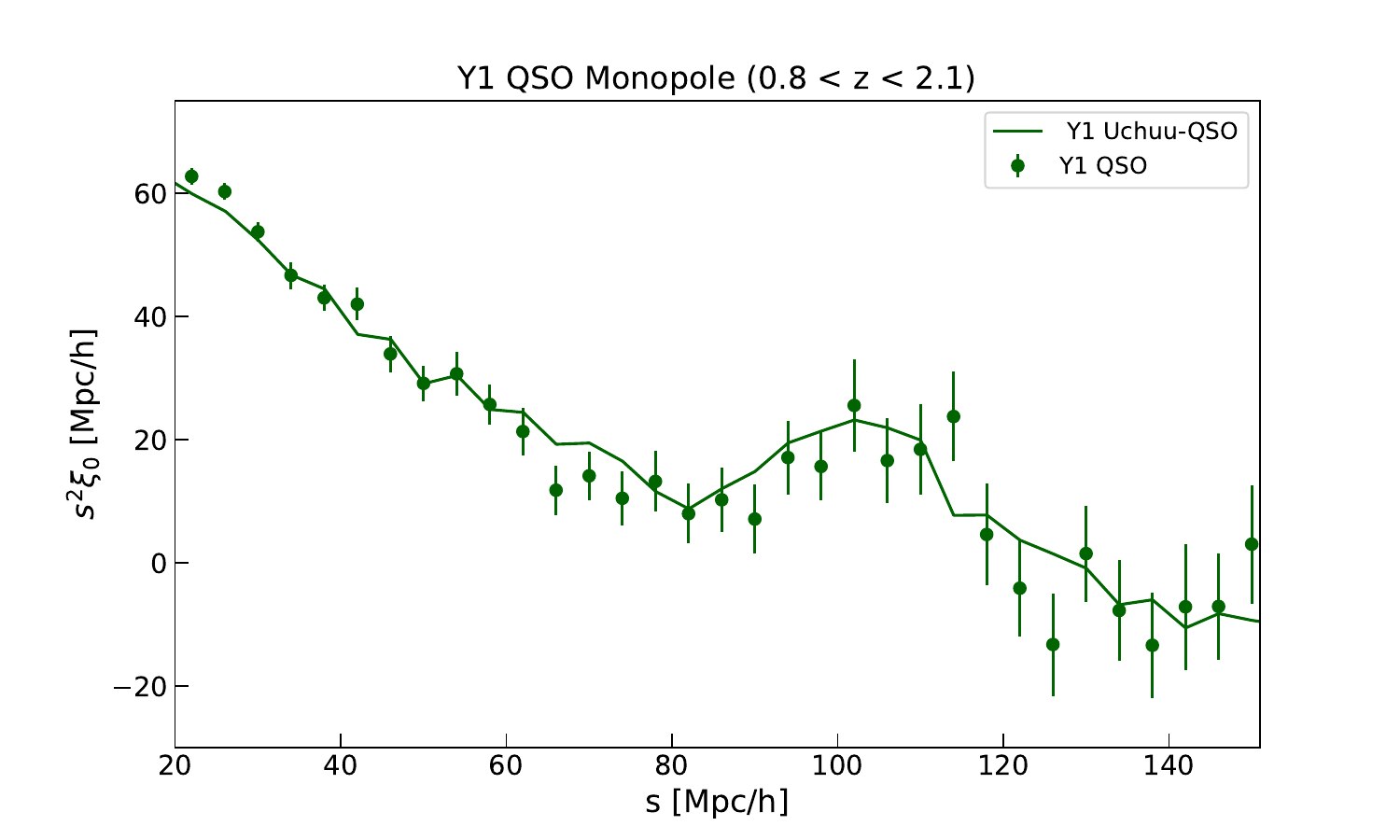}
    
    
    \caption{Measurements of the monopole of the redshift-space correlation function for $0.8 < z < 1.6$ (ELGs) and $0.8 < z < 2.1$ (QSOs), showing the BAO peak around $100 \hMpc$.  The points with error bars represent the measurements from the DESI DR1. 
    The results show agreement between theory and observation.}
    \label{fig:y1-2pcf}
\end{figure*}

\begin{figure*}

    \includegraphics[width=0.54\columnwidth]{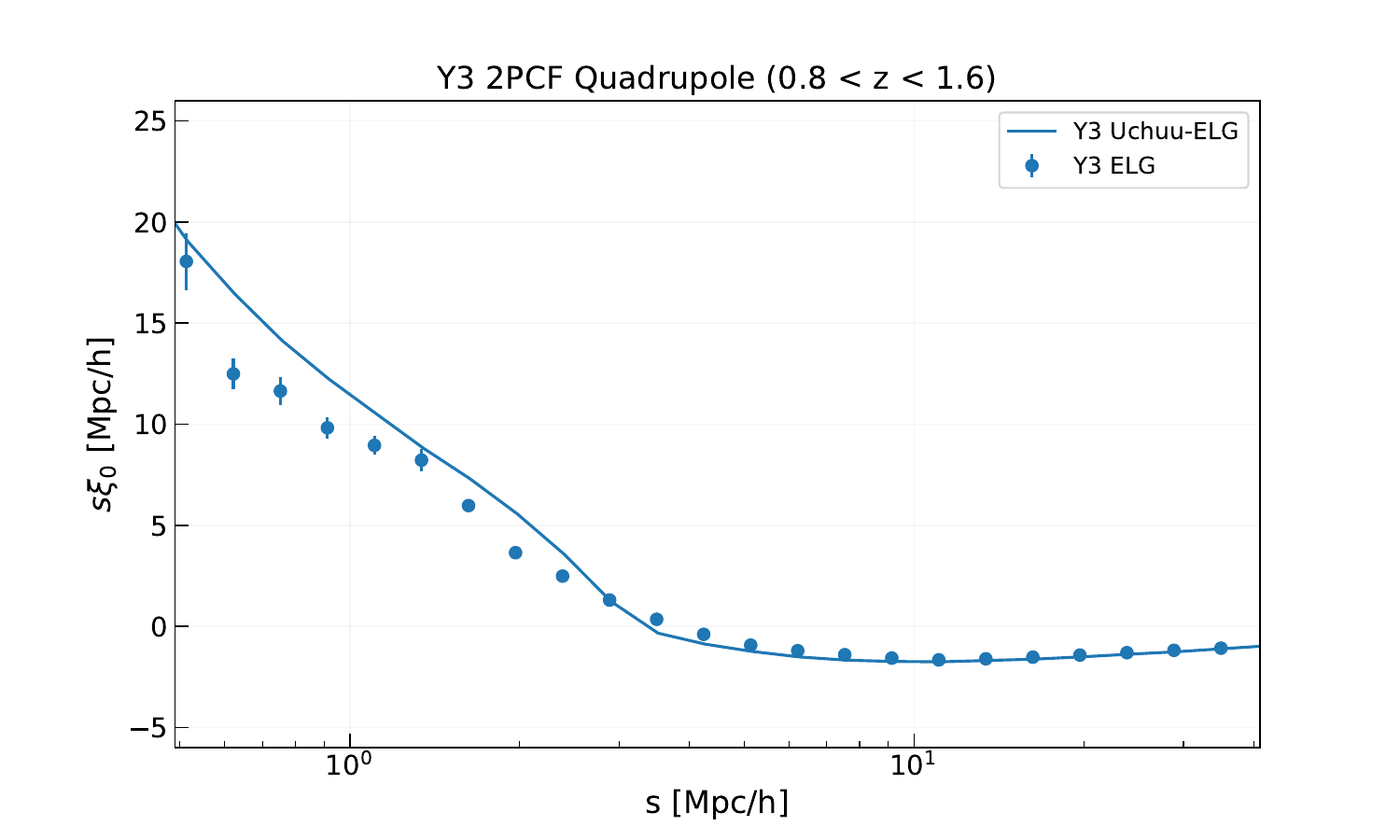}
    \includegraphics[width=0.54\columnwidth]{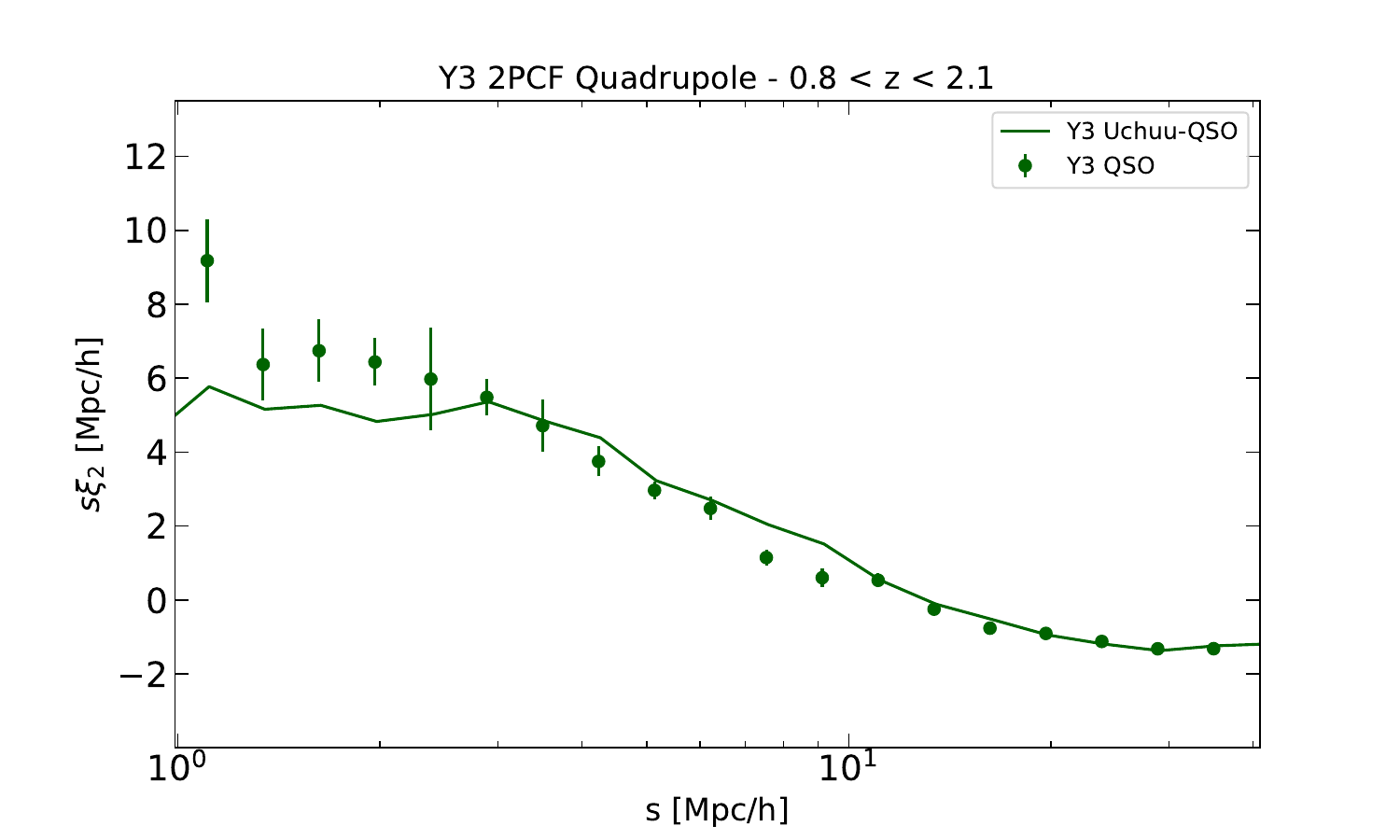}
    
    \caption{Measurements of the quadrupole of the redshift-space correlation function for $0.8 < z < 1.6$ (ELGs) and $0.8 < z < 2.1$ (QSOs).  The points with error bars represent the quadrupole measurements from the DESI DR2. Even though we only fit to the monopole, the quadrupole results show agreement between theory and observation. }
    \label{fig:y3-2pcf_quad}
\end{figure*}

Figures~\ref{fig:sv3-2pcf}, ~\ref{fig:y1-2pcf} \& ~\ref{fig:y3-2pcf_quad} present the measurements of the redshift-space correlation function. The monopole, $\xi_0(s)$, and quadrupole, $\xi_2(s)$, are shown for different redshift intervals indicated in the legends. The points with errors indicate the DESI clustering measurements, and the solid curves represent the theoretical predictions based on Planck cosmology, determined from the  \Uchuu DR2 lightcones, in the redshift intervals indicated in the figure caption, see also Table~\ref{tab:all-basic}. Overall, the data is consistent with theoretical predictions. 

\begin{figure*}

    \includegraphics[width=0.54\columnwidth]{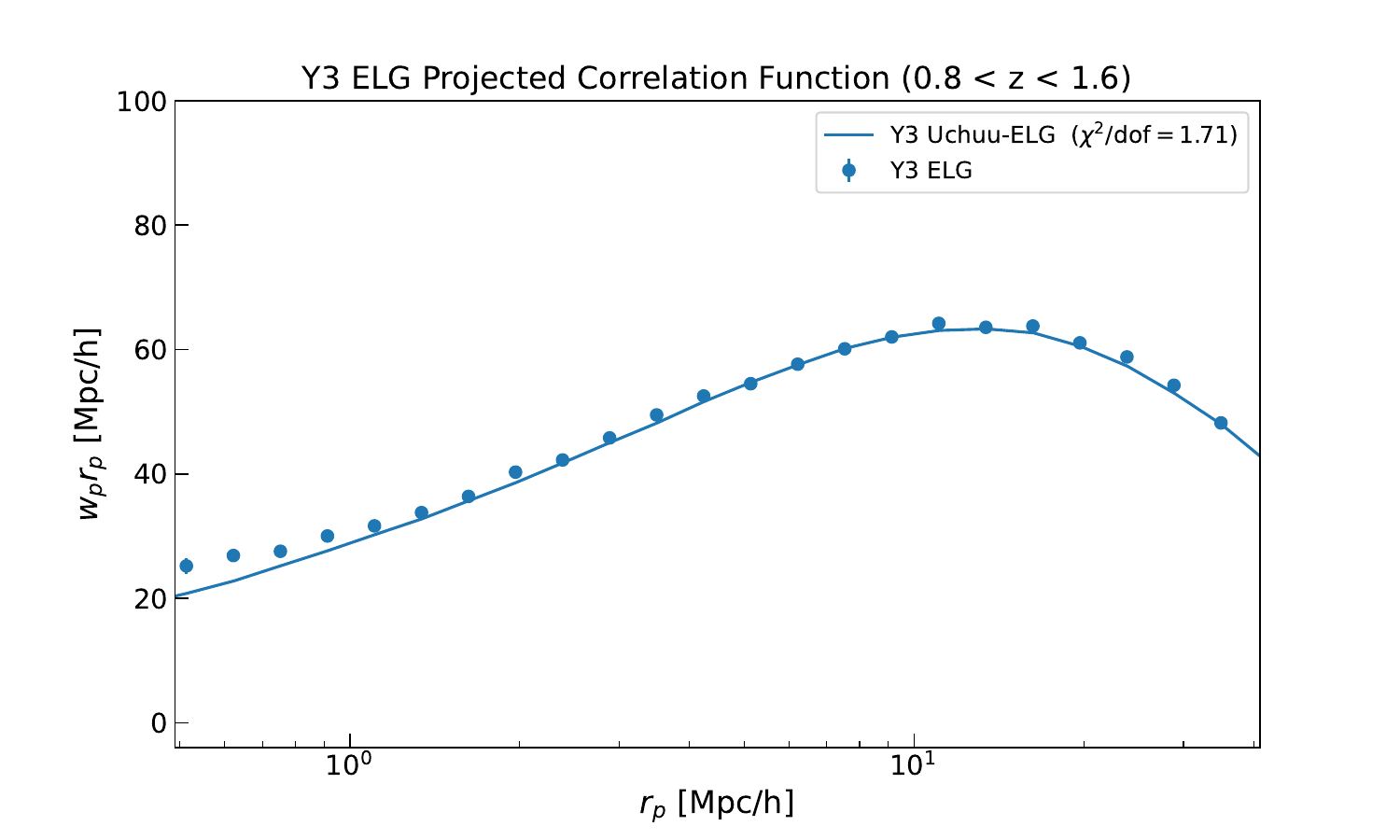}
    \includegraphics[width=0.54\columnwidth]{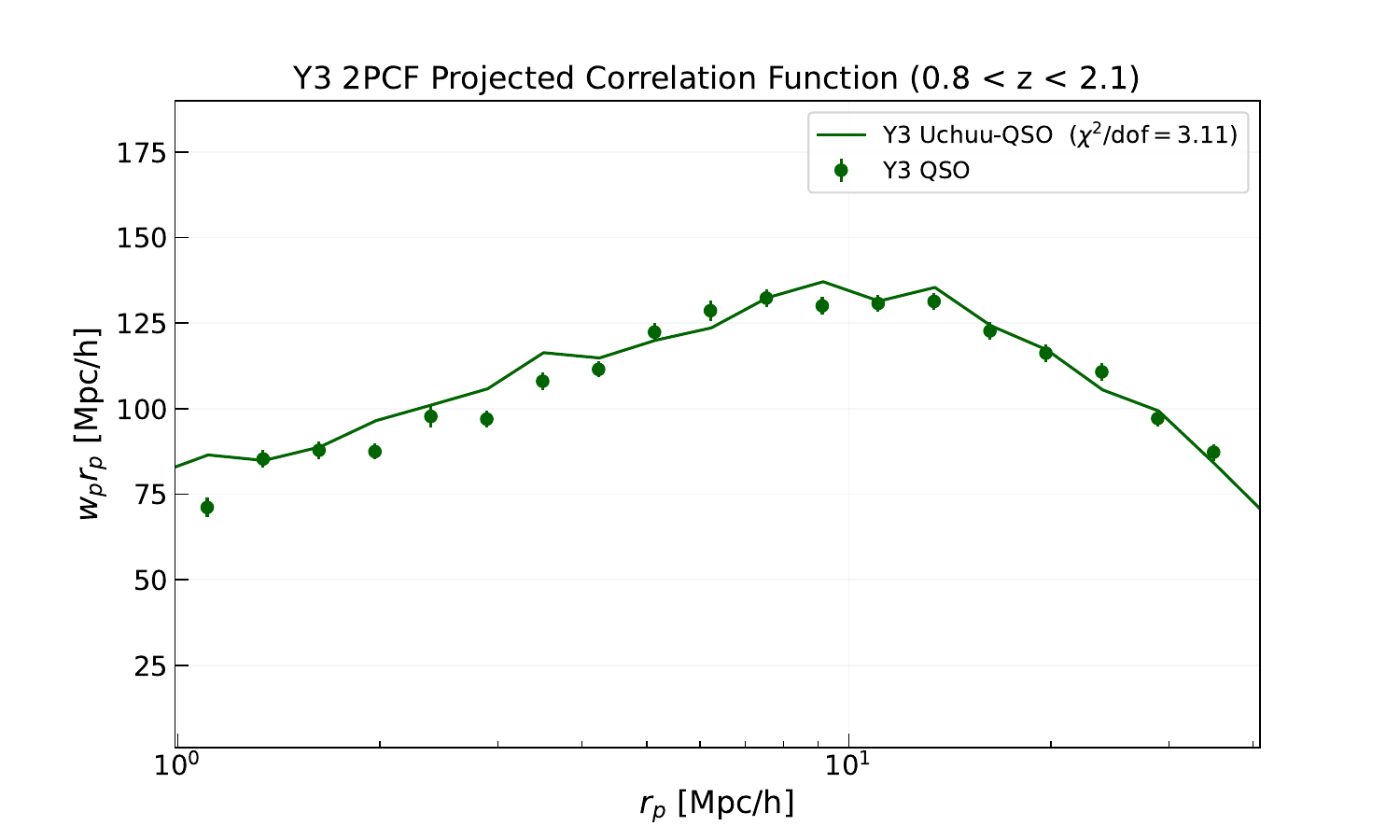}
    
    \caption{Measurements of the Projected correlation function for $0.8 < z < 1.6$ (ELGs) and $0.8 < z < 2.1$ (QSOs).  The points with error bars represent the projected correlation measurements from the DESI DR2. Even though we only fit to the monopole, the results here show agreement between theory and observation. }
    \label{fig:y3-2pcf_projected}
\end{figure*}

For the QSO sample, once we account for the redshift errors, we find good agreement of both the monopole and the quadrupole within the fit region ($3~\hMpc < s < 70~\hMpc$).

\subsubsection{Clustering dependence on redshift}

\begin{figure*}

    \includegraphics[width=0.325\columnwidth, height=4cm]{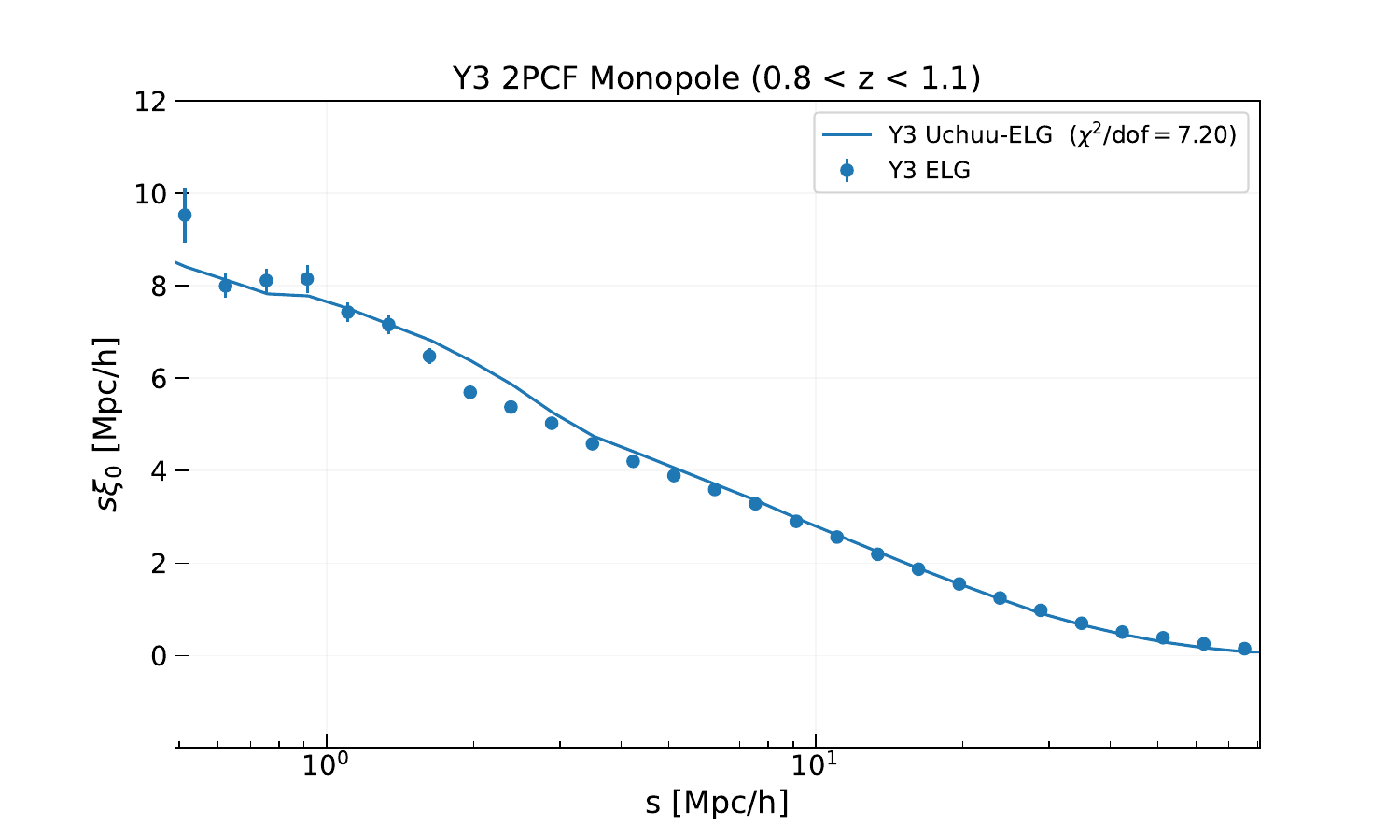}
    \includegraphics[width=0.325\columnwidth, height=4cm]{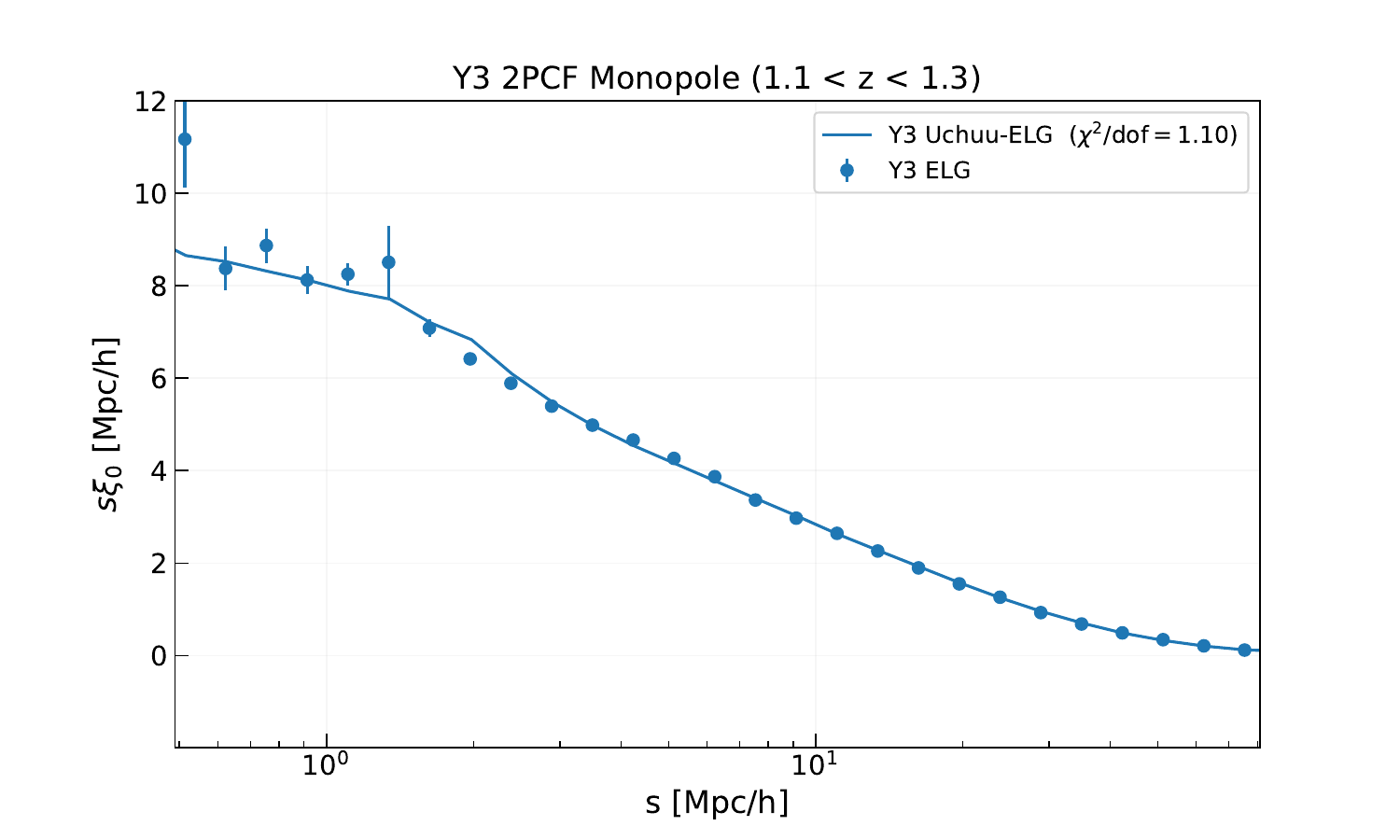}
    \includegraphics[width=0.325\columnwidth, height=4cm]{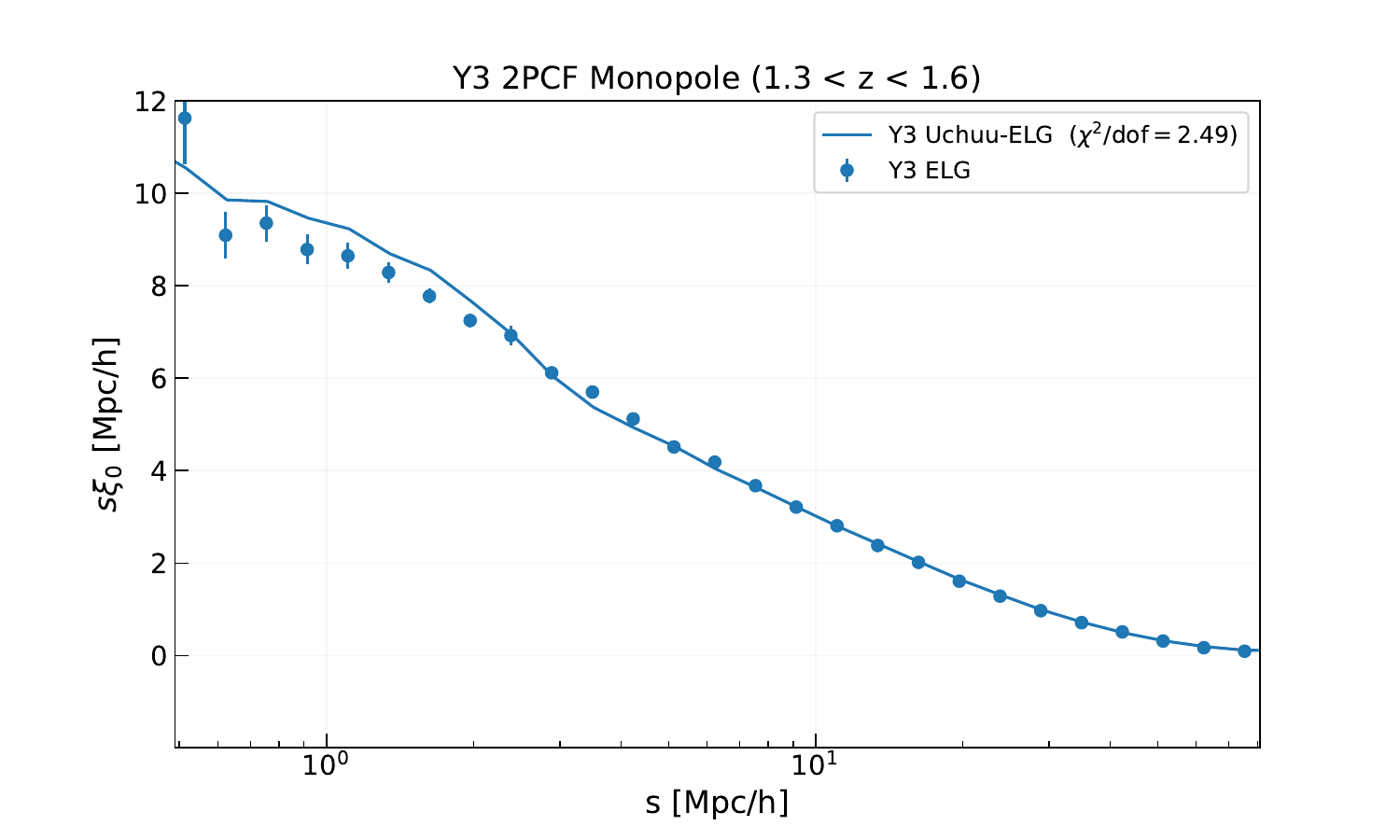}

    \caption{ELG clustering across different redshift bins: $0.8 < z < 1.1$, $1.1 < z < 1.3$, and $1.3 < z < 1.6$. The $V_\mathrm{mean}(z)$ model is able to account for redshift evolution reasonably. }
    \label{fig:sv3-2pcf-elg}
\end{figure*}

\begin{figure*}

    \includegraphics[width=0.325\columnwidth, height=4cm]{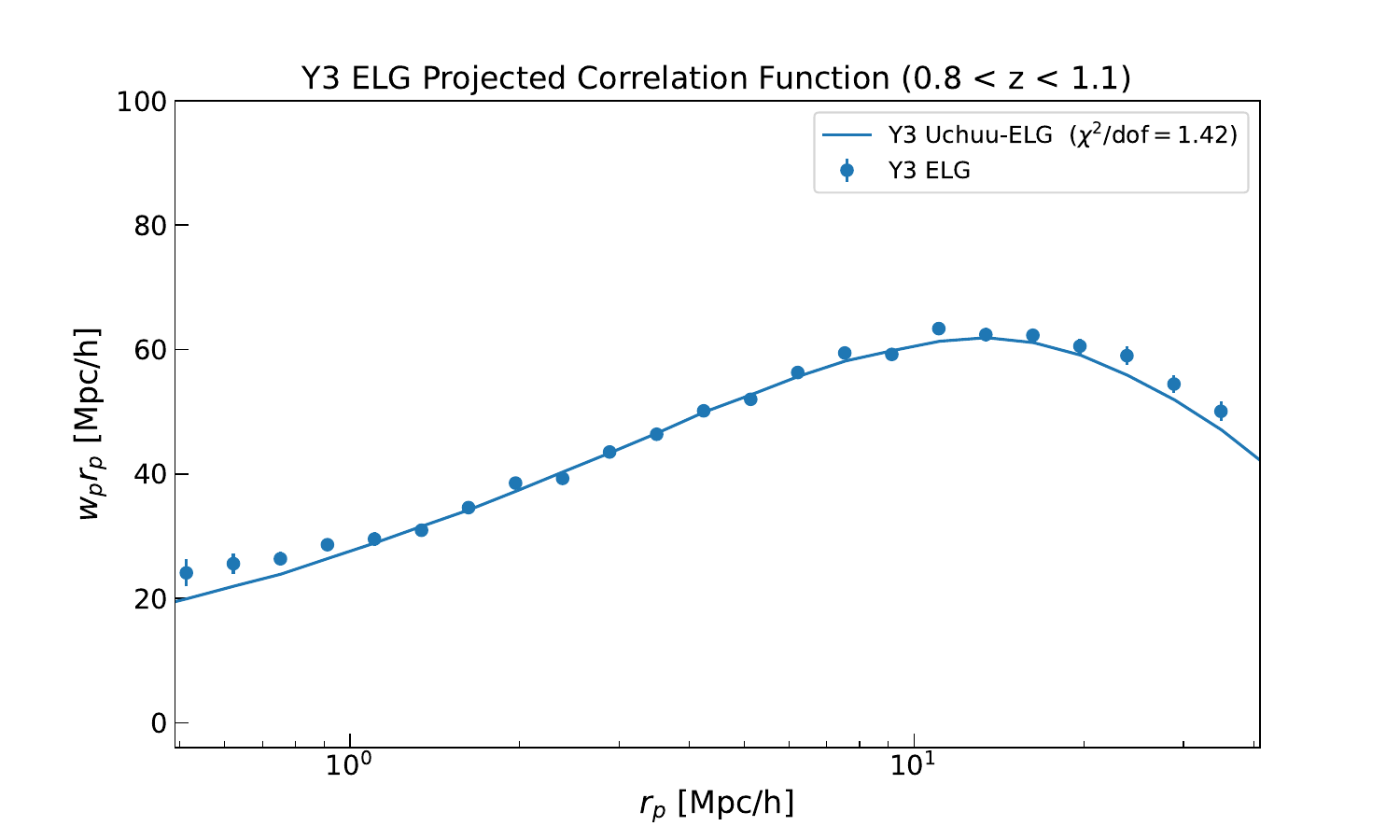}
    \includegraphics[width=0.325\columnwidth, height=4cm]{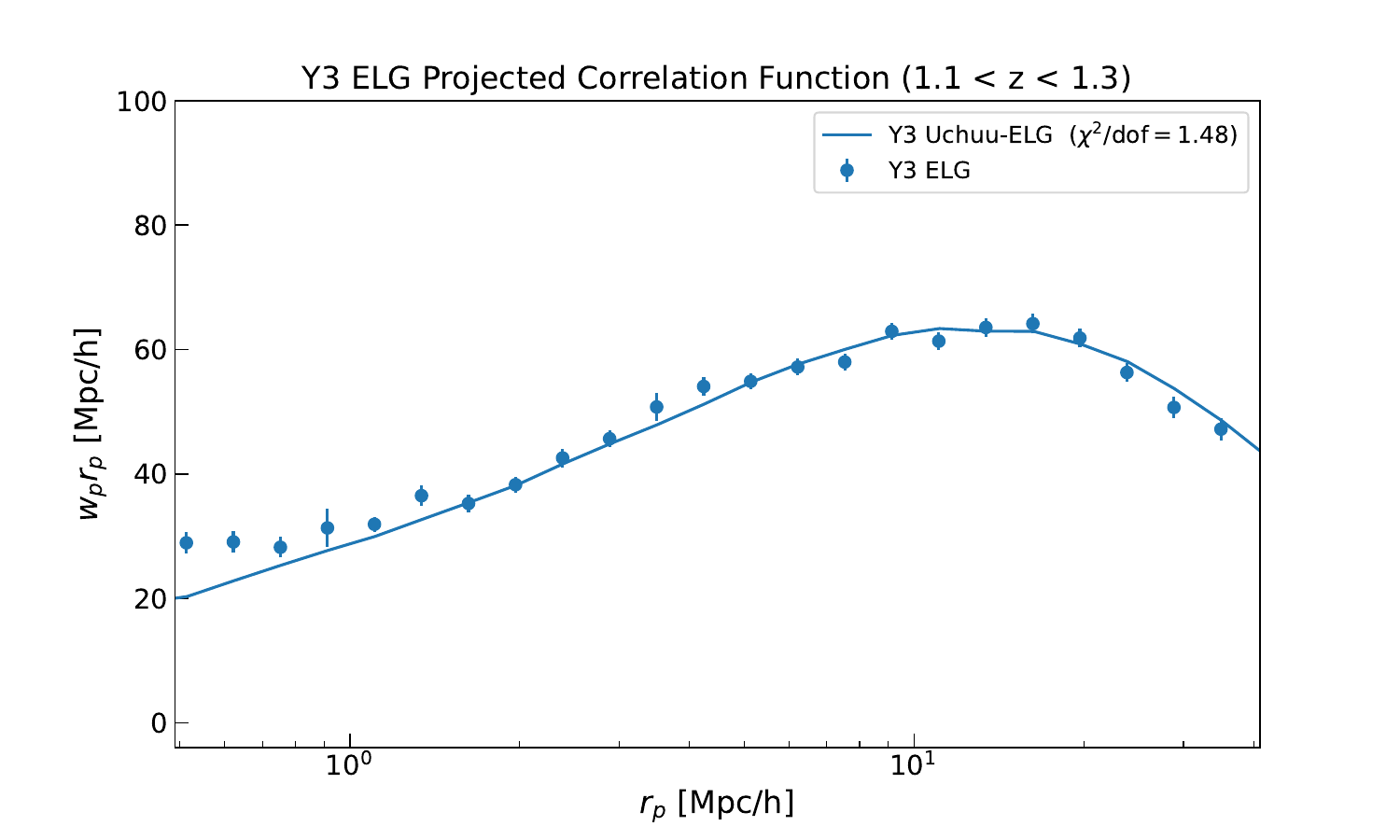}
    \includegraphics[width=0.325\columnwidth, height=4cm]{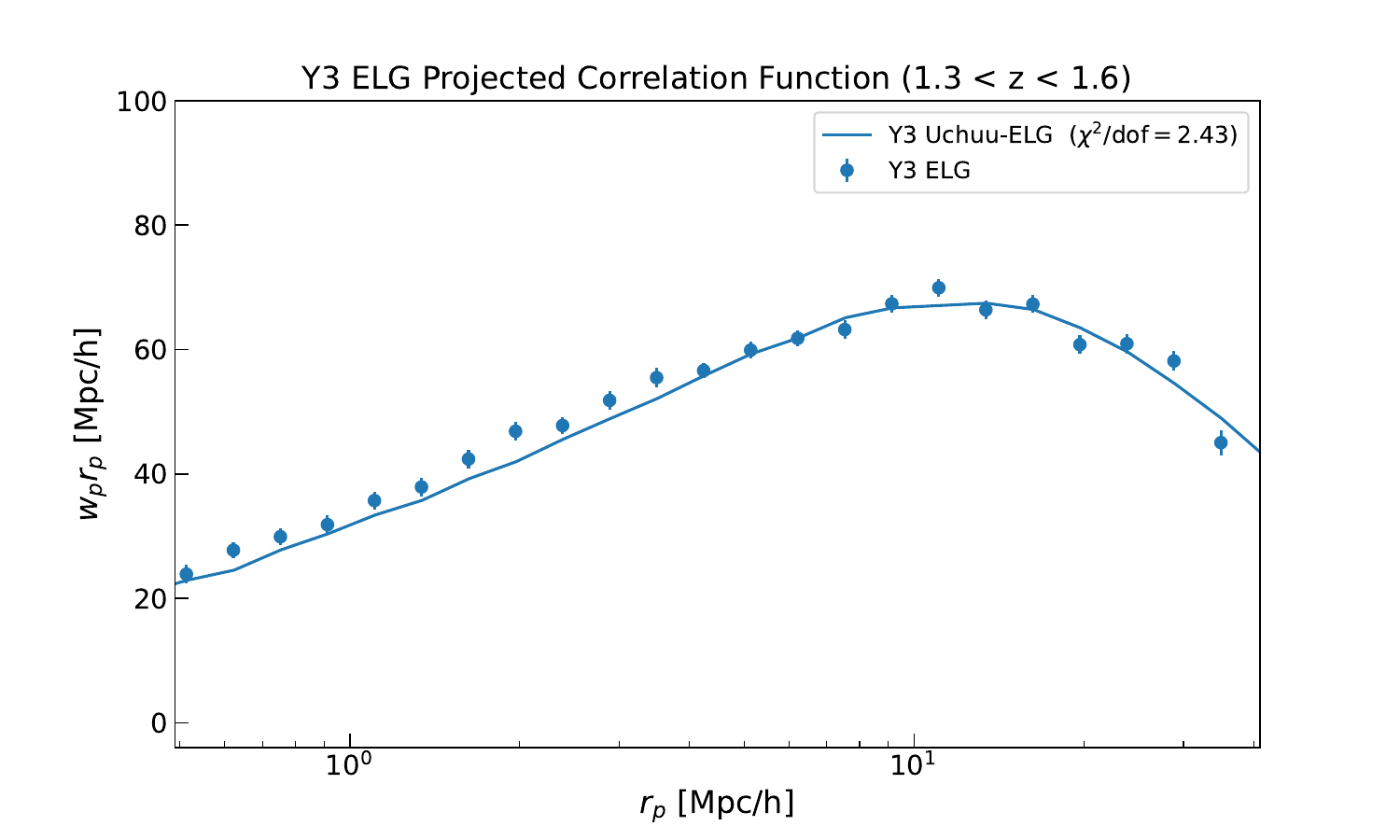}

    \caption{Projected correlation functions for ELG across the same bins as in Figure \ref{fig:sv3-2pcf-elg}.}
    \label{fig:sv3-2pcf-elg-projected}
\end{figure*}

The correlation functions of the ELGs, split into narrower redshift bins, are shown in Figure~\ref{fig:sv3-2pcf-elg} and Figure~\ref{fig:sv3-2pcf-elg-projected}. The apparent visual agreement of the ELG monopole is better in all panels of Figure~\ref{fig:sv3-2pcf-elg} compared to Figure~\ref{fig:sv3-2pcf}, due to the reduced separation range appropriate to the binned data. 

\begin{figure*}

    \includegraphics[width=0.525\columnwidth]{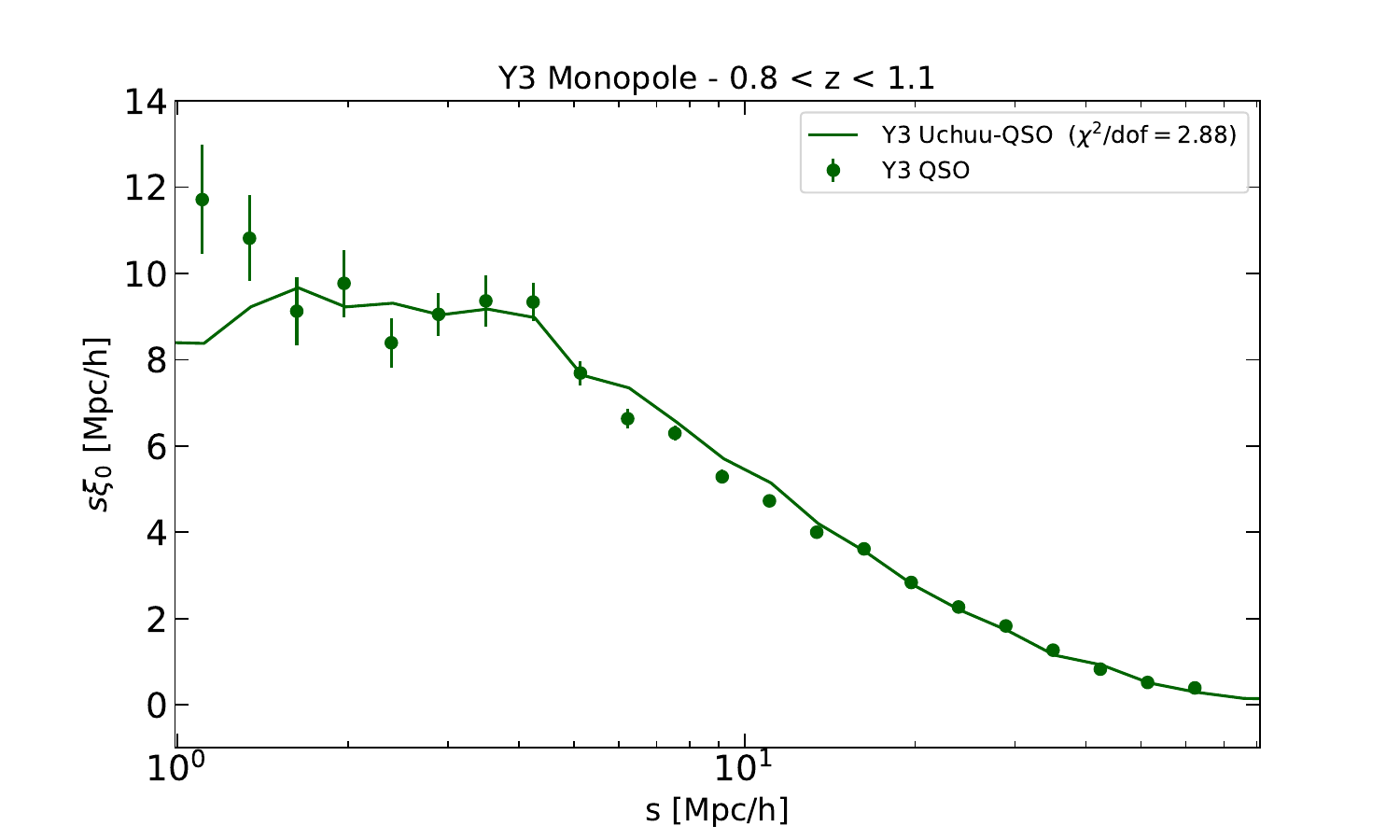}
    \includegraphics[width=0.525\columnwidth]{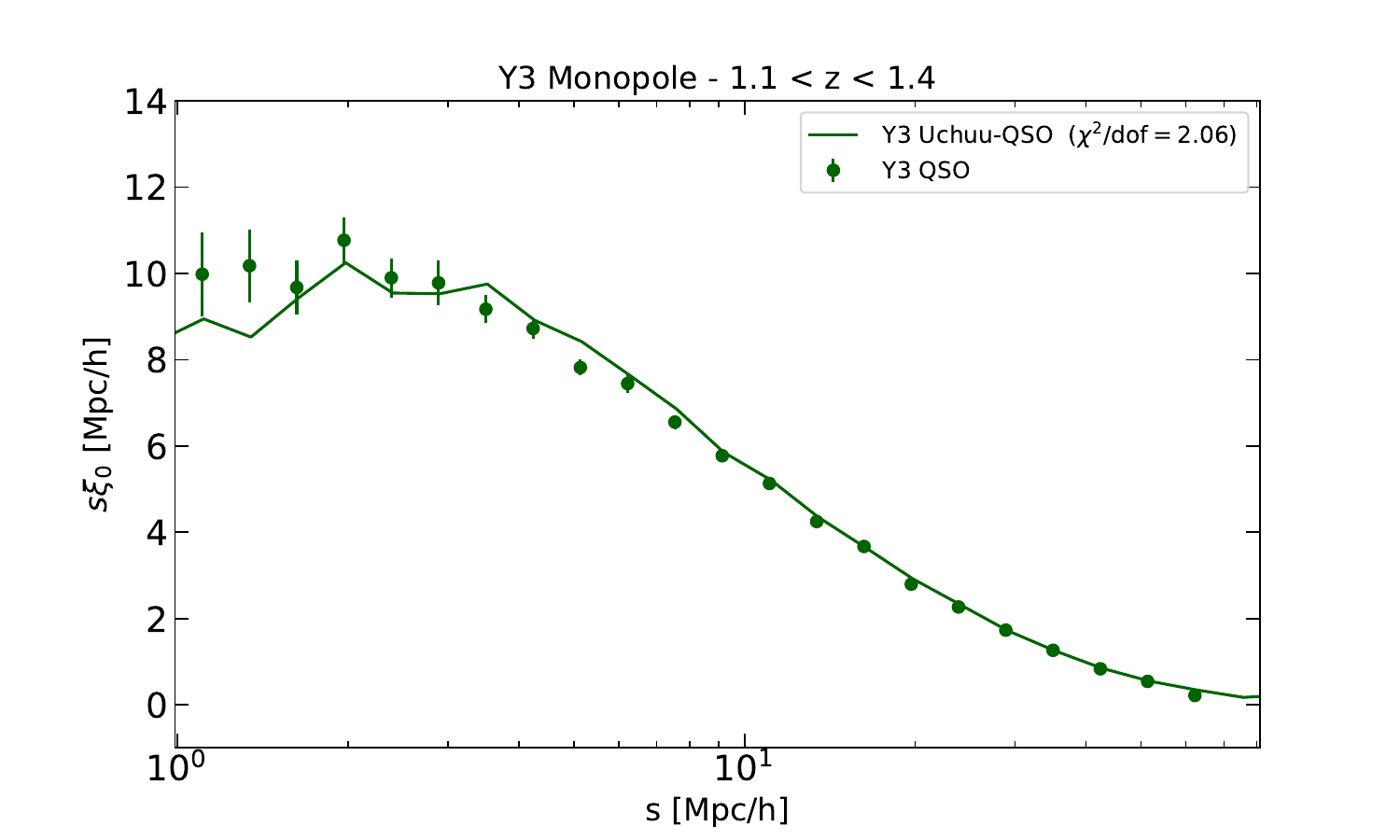}
    \includegraphics[width=0.525\columnwidth]{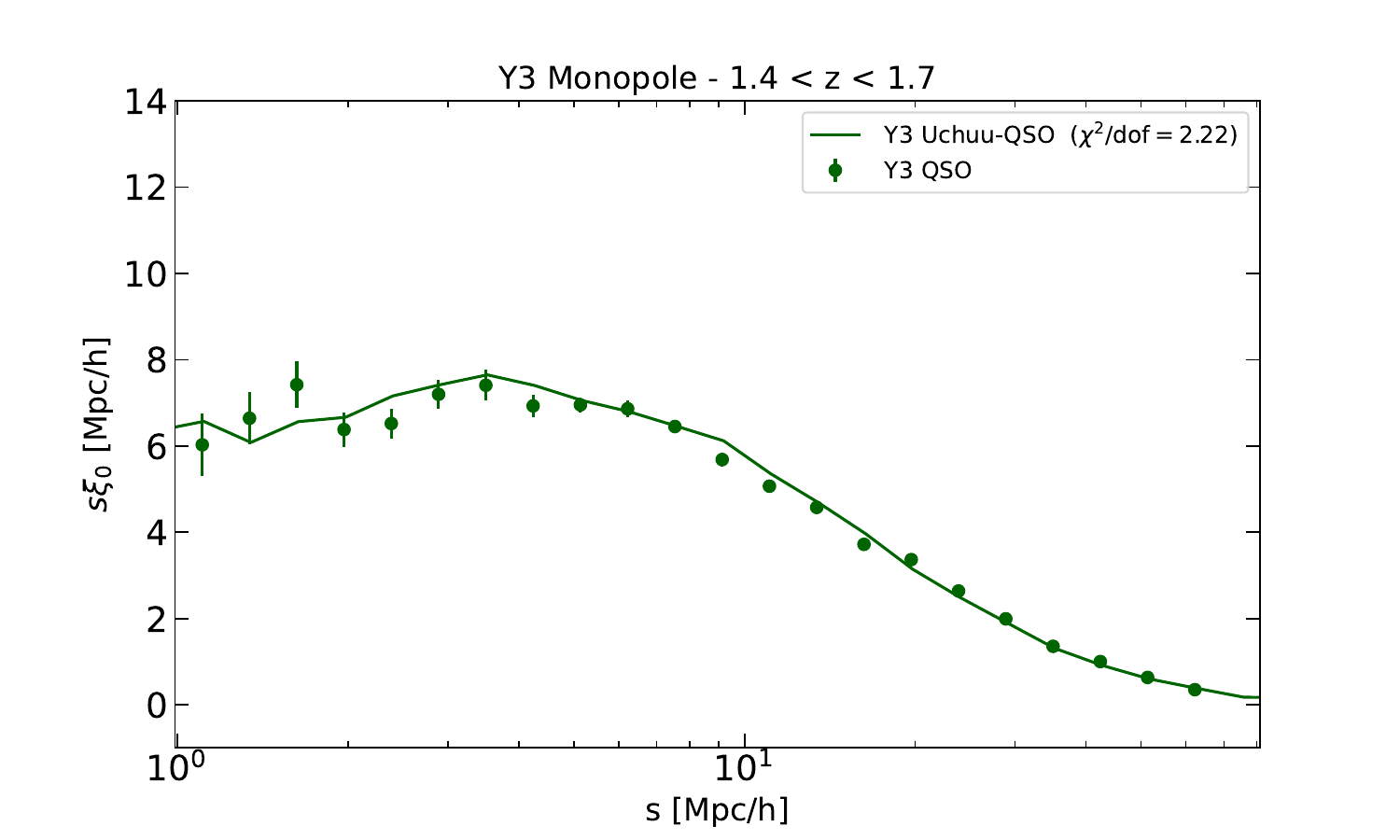}
    \includegraphics[width=0.525\columnwidth]{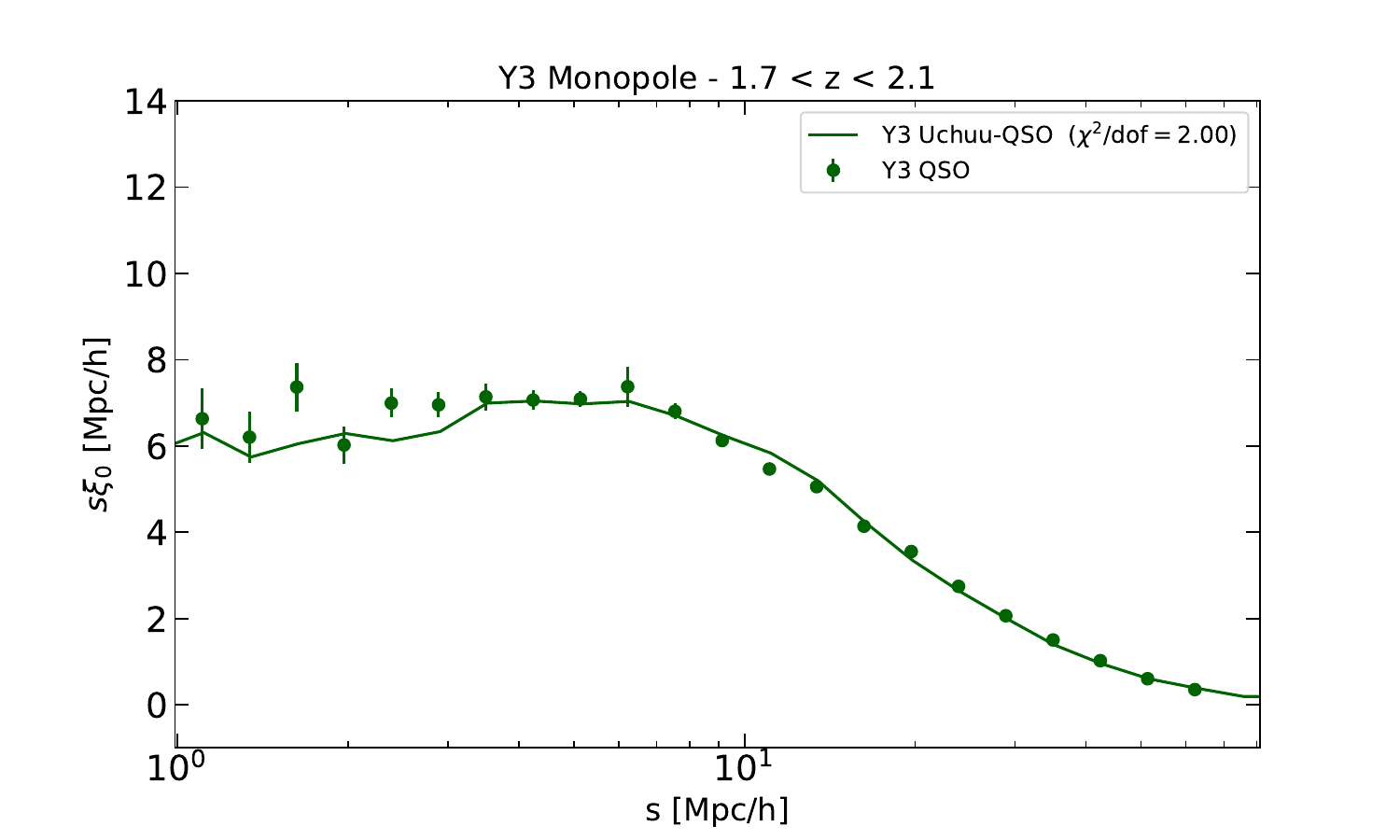}
    \caption{QSO clustering across different redshift bins: $0.8 < z < 1.1$, $1.1 < z < 1.4$, $1.4 < z < 1.7$, and $1.7 < z < 2.1$.  The $f_\mathrm{sat}(z)$ model is able to account for redshift evolution reasonably.}
    \label{fig:sv3-2pcf-qso}
\end{figure*}


\begin{figure*}

    \includegraphics[width=0.5\columnwidth]{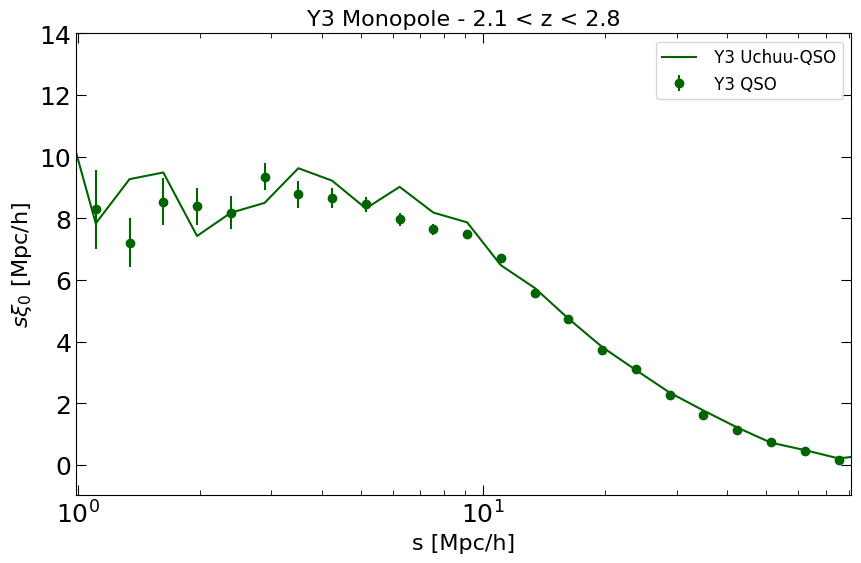}
    \includegraphics[width=0.5\columnwidth]{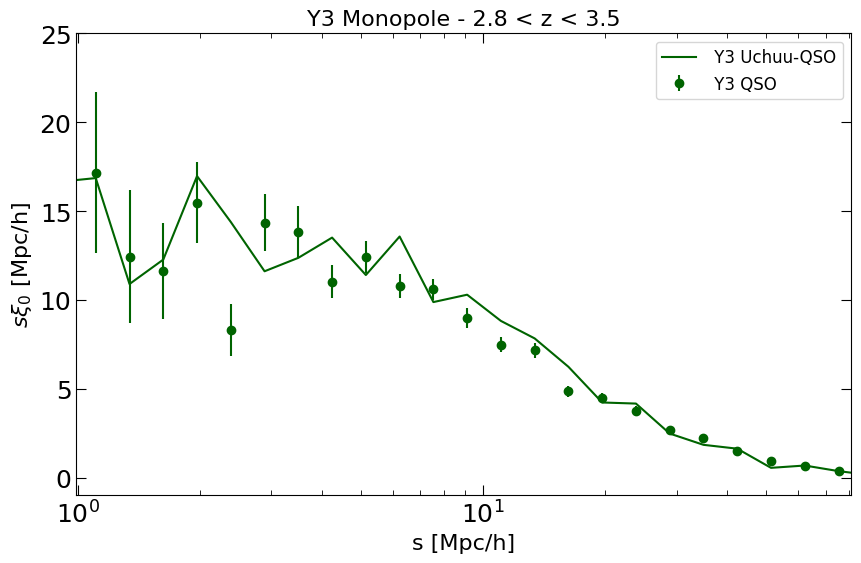}
    \caption{QSO clustering across  higher redshift bins - $2.1 < z < 2.8$ and $2.8 < z < 3.5$, where we simply extended our best-fit model from $0.8 < z < 2.1$. }
    \label{fig:sv3-2pcf-qso-highz}
\end{figure*}

\begin{figure*}

    \includegraphics[width=0.525\columnwidth]{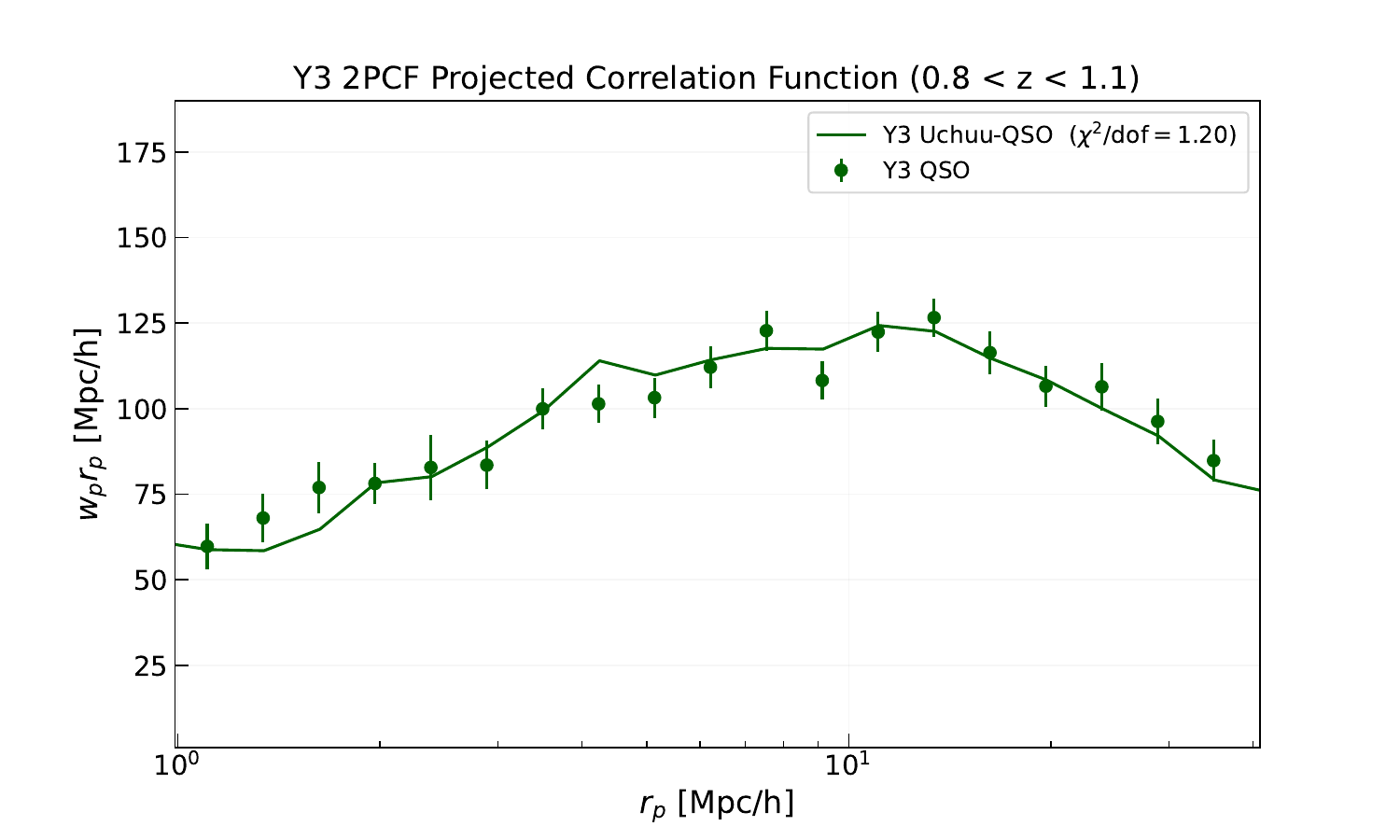}
    \includegraphics[width=0.525\columnwidth]{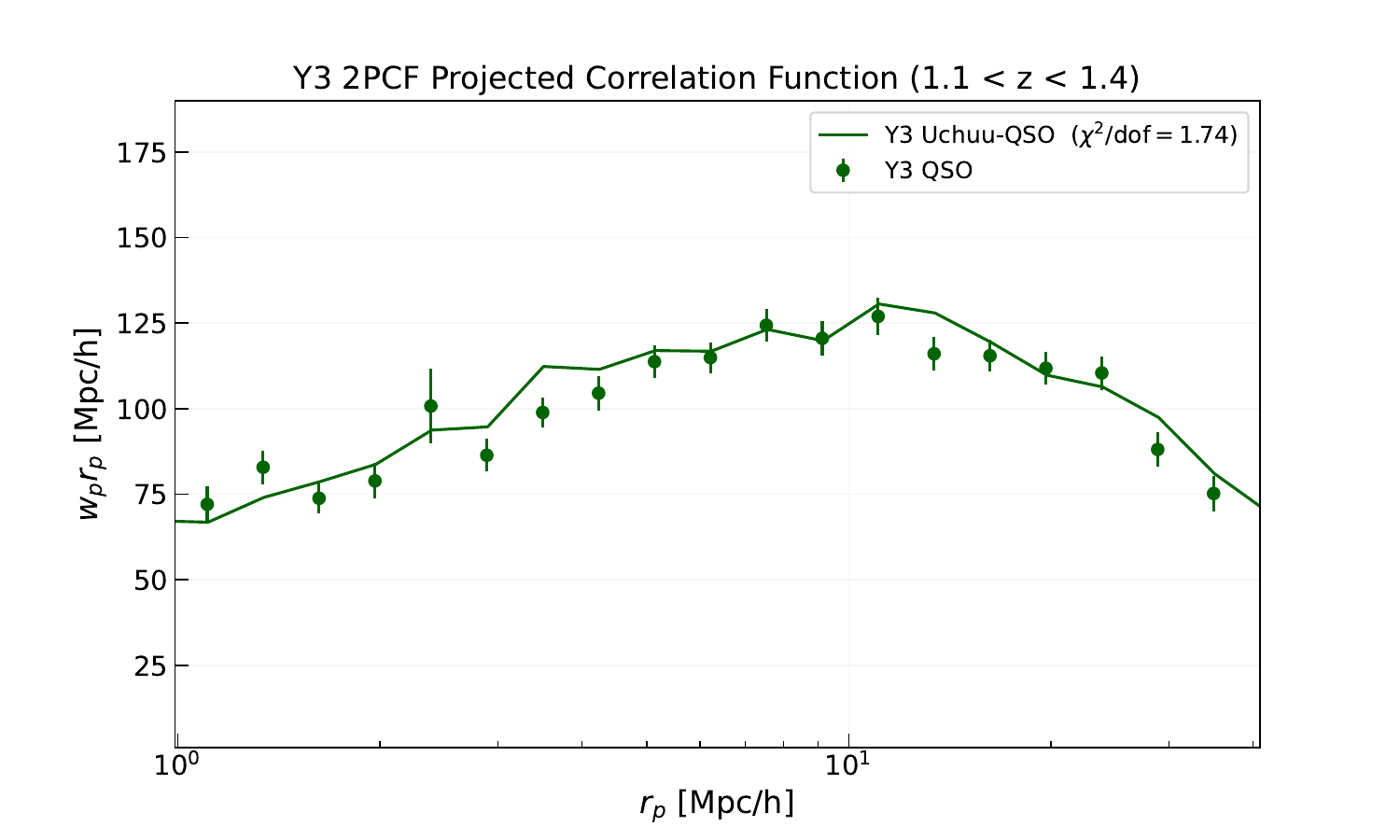}
    \includegraphics[width=0.525\columnwidth]{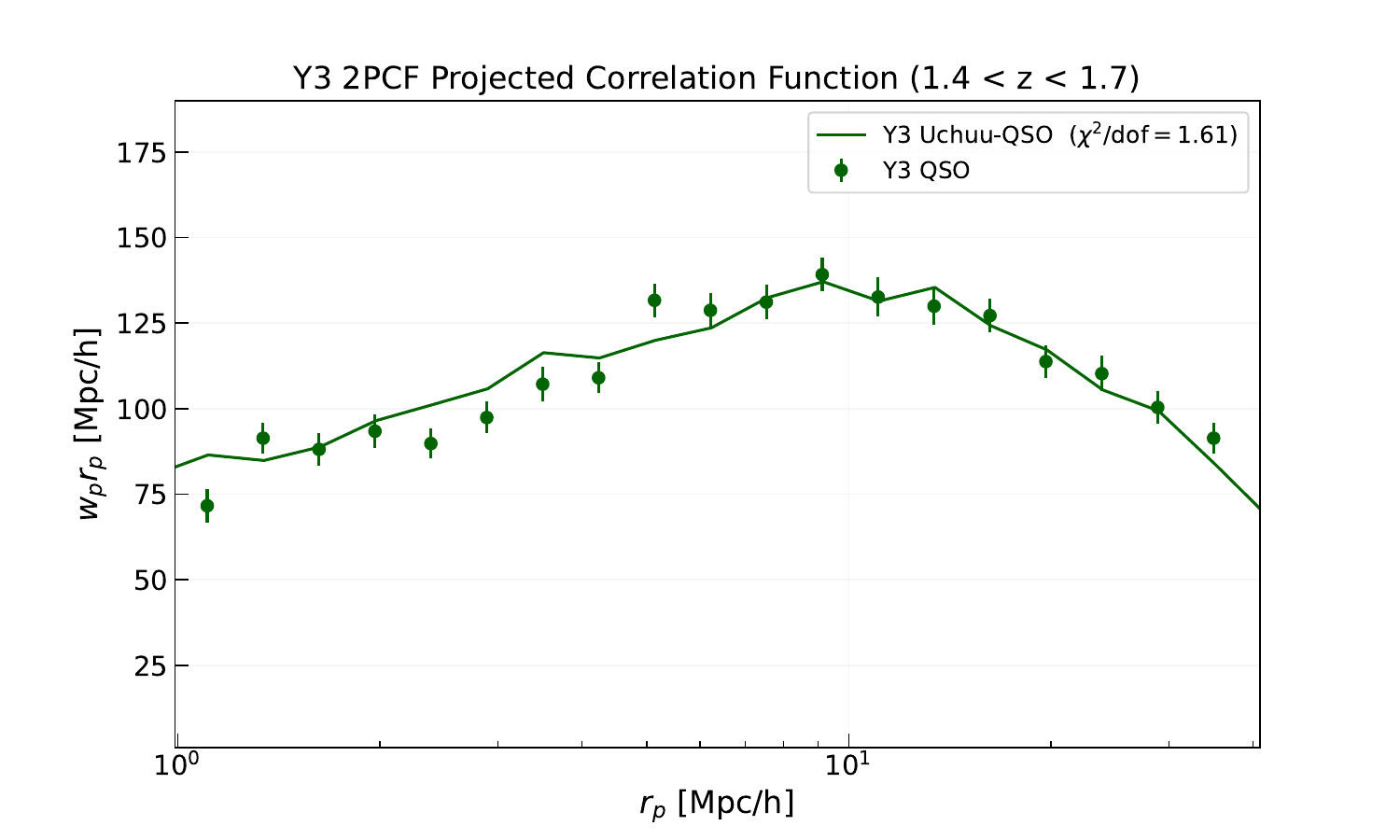}
    \includegraphics[width=0.525\columnwidth]{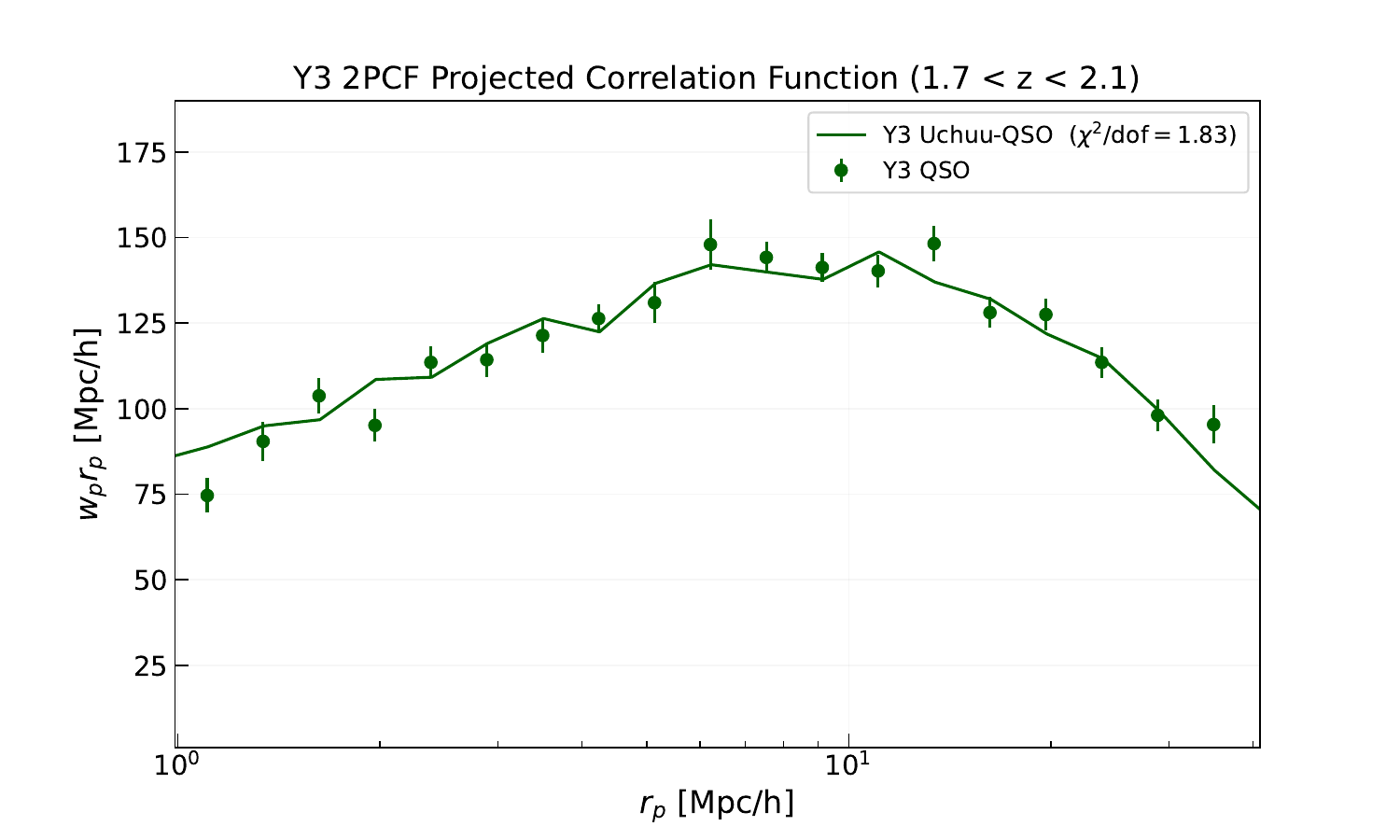}
    \caption{Projected Correlation functions for QSO across same redshift bins as in Figure~\ref{fig:sv3-2pcf-qso}.  }
    \label{fig:sv3-2pcf-qso-projected}
\end{figure*}

\noindent For QSOs, Figures~\ref{fig:sv3-2pcf-qso},~\ref{fig:sv3-2pcf-qso-highz} and~\ref{fig:sv3-2pcf-qso-projected} show the correlation function in redshift bins. We find good agreement between \Uchuu and the DESI DR2 data in the respective bins.    

\begin{figure*}
    \centering
    \includegraphics[width=0.48\textwidth]{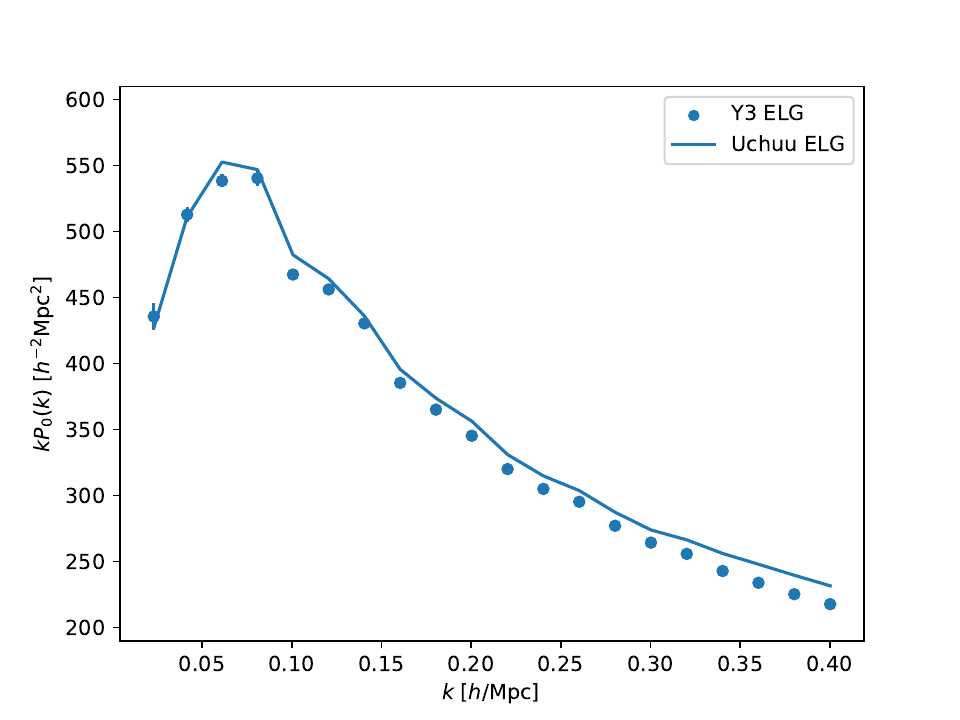}
    \hspace{0.02\textwidth}
    \includegraphics[width=0.48\textwidth]{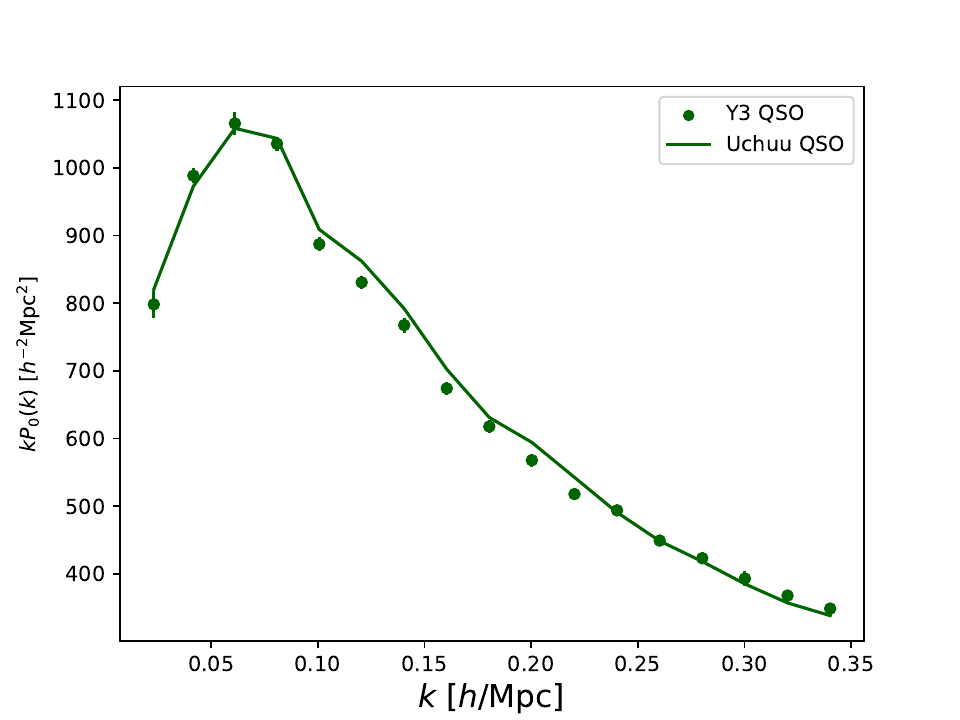}
    \caption{Power spectrum monopole measurements, for ELGs (left panel) and QSOs (right panel). The dotted lines represent Y3 data, and the dashed lines are the measurements on the \Uchuu mocks. Since we do not fit to the power spectrum, there are subtle deviations at high $k$ (in ELGs), yet the model recovers reasonable clustering at most scales. }
    \label{fig:sv3-power-spectrum}
\end{figure*}

\subsubsection{Redshift Space Power spectrum}

We also measure the power spectrum monopole, $P_0(k)$, and quadrupole, $P_2(k)$, with the Python package \textsc{pypower}\footnote{\url{https://github.com/cosmodesi/pypower/}} which is based on the estimator from \citep{Hand2017}. Similarly as for the correlation function measurements, incompleteness weights are applied to the survey data, and for both the survey data and \Uchuu lightcones, FKP weights calculated from $n(z)$ and the same fiducial power $P_0$ as above are also applied to each tracer. To minimise the amount of aliasing from discrete Fourier sampling, we have used the piecewise cubic spline (PCS) mesh assignment scheme with a grid number $N_\mathrm{grid} = 1024$ in each dimension with interlacing~\citep{Sefusatti2016}. For each of the two tracers,  Figure~\ref{fig:sv3-power-spectrum} shows the power spectrum multipoles over the wavenumber range $k \in [0.005, 0.405]~h\mathrm{Mpc}^{-1}$ in 20 uniform bins.

The power spectrum measurements performed here are similar to the BOSS and eBOSS power spectra~\citep{Beutler2017,GilMarin2020,mattia2021,Neveux2020}, where the local plane-parallel approximation is adopted to account for a varying line of sight~\citep{Feldman1994,Yamamoto2006}. The local line of sight is chosen to be the end-point vector to one of the galaxies in a pair, which enables fast, FFT-based evaluations to be carried out~\citep{Bianchi2015}. A minor difference here is that the normalization factor is computed directly from the mesh field instead of relying on the angularly uniform quantity,~\( n(z) \); by using both the data and random mesh fields sampled with cell size $10~h^{-1}\,\mathrm{Mpc}$, the normalization factor approximates the window function amplitude and thus the power spectrum amplitude can be approximately compared across different survey geometries~\citep{2024arXiv241112020D}. 

\subsection{Mean Halo-Occupancy}

\begin{figure*}
    \centering
    
    \includegraphics[width=0.48\columnwidth]{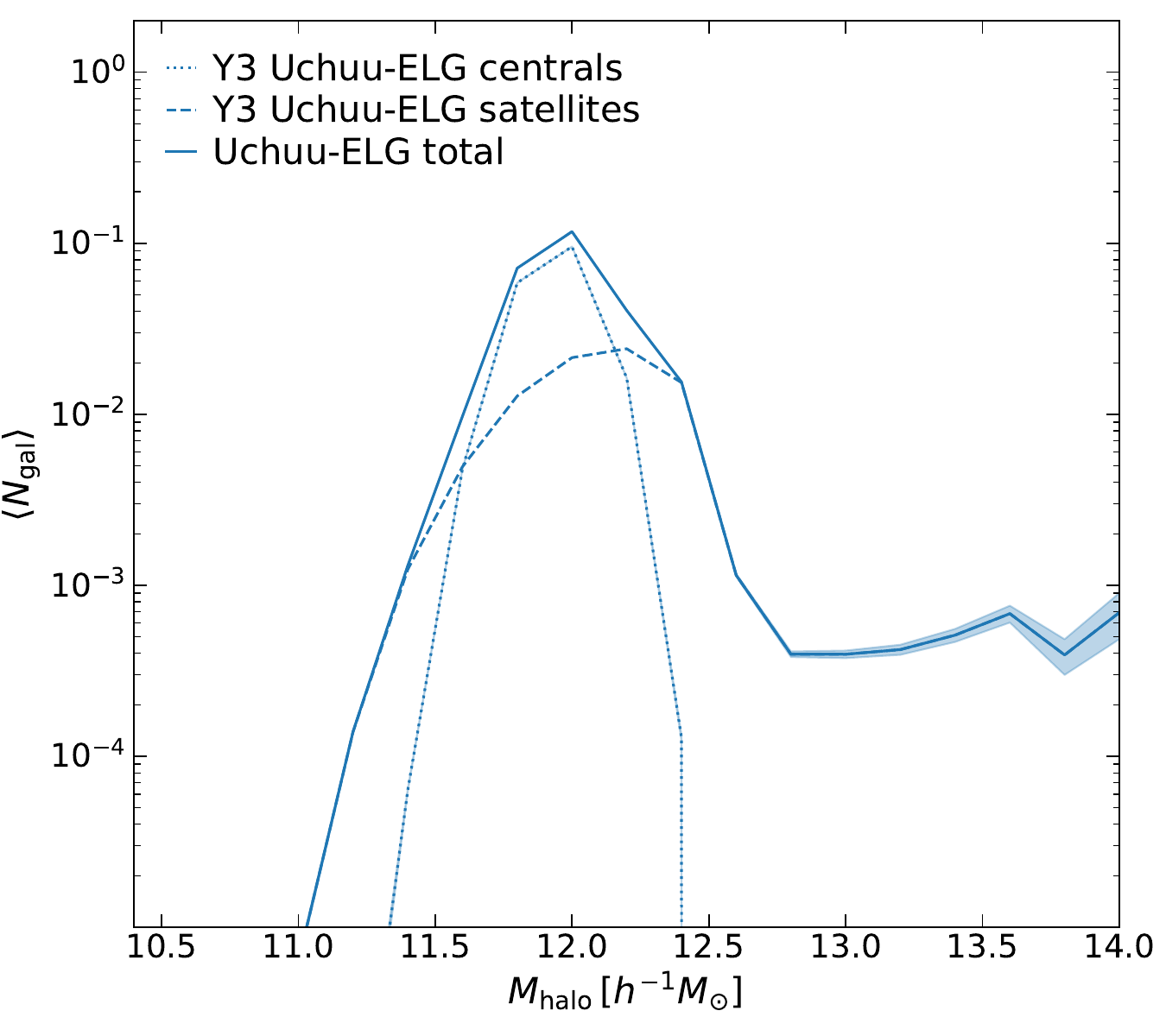}
    \hspace{0.02\textwidth}
    \includegraphics[width=0.48\columnwidth]{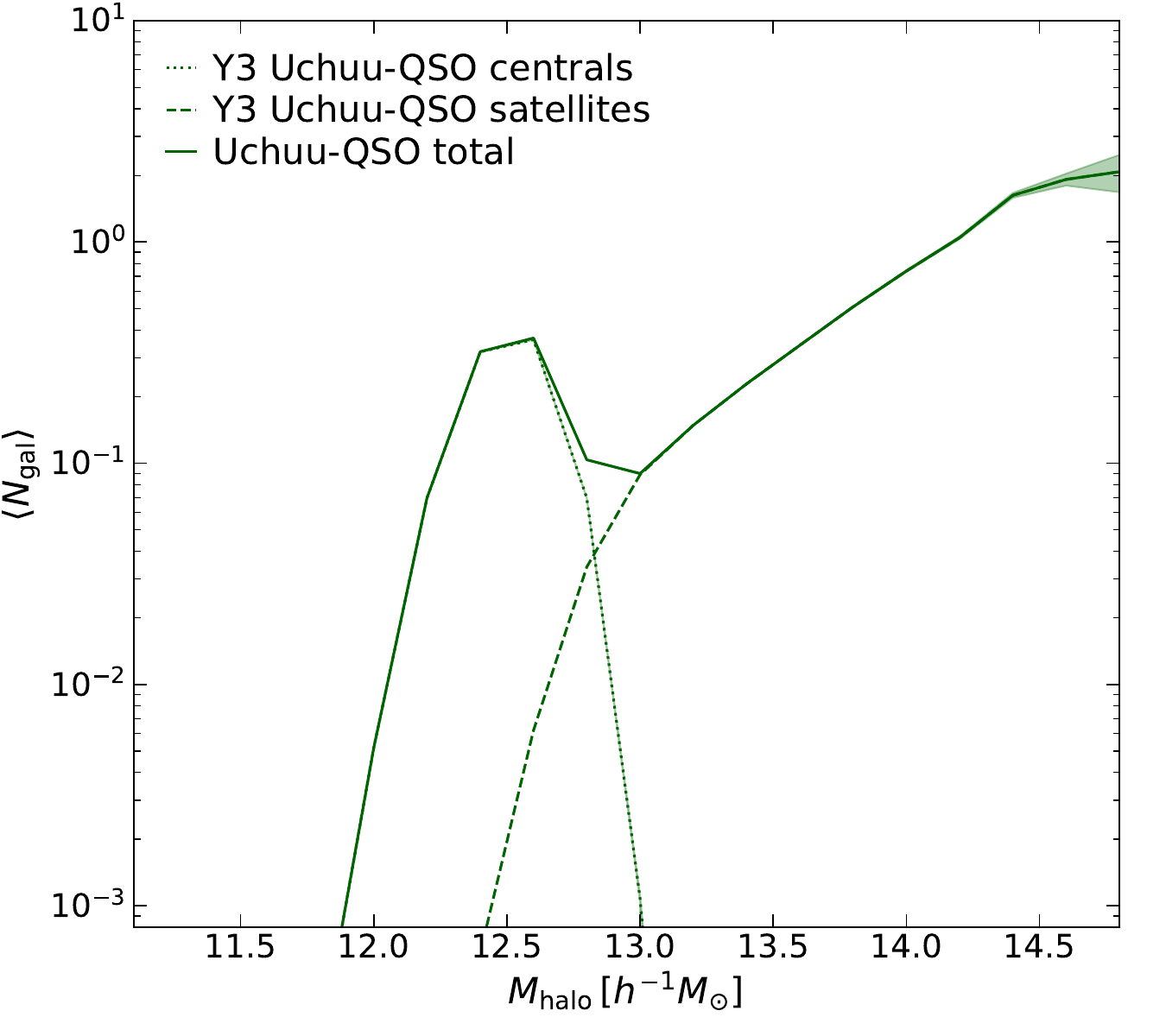}
    \caption{Mean halo occupancy of ELGs (left) and QSOs (right) as determined from our SHAM Uchuu boxes. The redshift of the Uchuu-ELG box is $z = 1.22$, while the redshift of
the Uchuu-QSO box is also $z = 1.22$. The mean number of galaxies of a halo with a given mass $ M_\mathrm{halo}$ is denoted by $< N_\mathrm{gal} >$.
The central galaxies are shown
with dotted lines and satellites with dashed lines. The solid lines represent the combined centrals and satellite occupation. The shaded area indicates the 1$\sigma$ uncertainty of the occupation measured from the Uchuu lightcone.}
    \label{fig:all-hod}
\end{figure*}

The abundance matching technique implemented in \Uchuu provides a complete determination of the distribution and properties of DESI tracers within their host dark matter haloes. This allows us to estimate the mean number of galaxies or quasars within a dark matter (sub)halo of virial mass $M_\mathrm{halo}$ for each tracer sample. In Figure~\ref{fig:all-hod}, we present the mean halo occupancy as a function of halo mass for ELG and QSO tracers, obtained from our independent set of \Uchuu lightcones. The clustering signal of the same samples for the DR2 is shown in Figure~\ref{fig:sv3-2pcf}). In our simulation, we can distinguish between central galaxies/quasars residing in their host haloes and satellite galaxies/quasars that live in subhaloes. By doing so, we are able to measure the HOD separately for centrals and satellites. These are shown by the dotted and dashed curves in Figure~\ref{fig:all-hod}, respectively, for each tracer.

For both ELG and QSO samples, the central HOD strongly shows the influence of our model with the occupation fraction resembling a Gaussian in $M_{\rm halo}$. This is expected due to the strong correspondence between $V_{\rm peak}$ and $M_{\rm halo}$. The satellite component rises more quickly at high $M_{\rm halo}$ for ELGs than QSOs. This is a result of the ELG sample having a lower best-fit $V_{\rm mean}$ than the QSO sample as shown in Table~\ref{tab:qso-shamparm} as well as the much larger sample size of ELGs. 


The HOD analysis of the ELG mocks reveals two mostly distinct components. The component of the HOD from the centrals follows a Gaussian distribution which dominates where $M_{\rm halo} < 10^{12.1}~\hMsun$, while the satellites exhibit a power-law distribution which dominates at  $M_{\rm halo} > 10^{12.4}~\hMsun$. The central haloes reach a peak occupation of 0.095, observed at $M_{\rm halo}=10^{12}~\hMsun$. 


Similar to the ELG HOD, the QSO HOD also exhibits a Gaussian distribution for the centrals, characterized by $M_{\rm halo} < 10^{12.7}~\hMsun$, and a power-law behavior for the satellites, which dominates at $M_{\rm halo} > 10^{12.8}~\hMsun$. The mean halo mass for the central quasars is $M_{\rm halo} = 10^{12.4}~\hMsun$, at peak occupancy of 0.319. This slightly deviates from the reported mean halo mass values reported in \citep{rodrigueztorres17} for eBOSS QSOs, ranging from $10^{12.5}~\hMsun$ to
 $10^{12.8}~\hMsun$. This is expected since the reported $V_\mathrm{mean}$ values in their study are higher compared to the values presented in this paper. However, despite these differences in numerical values, the shapes of the HOD in both DESI and eBOSS studies exhibit a similar pattern.

\subsection{Large Scale Bias}
\label{subsec:bias}
\begin{figure*}

    \includegraphics[width=0.49\columnwidth]{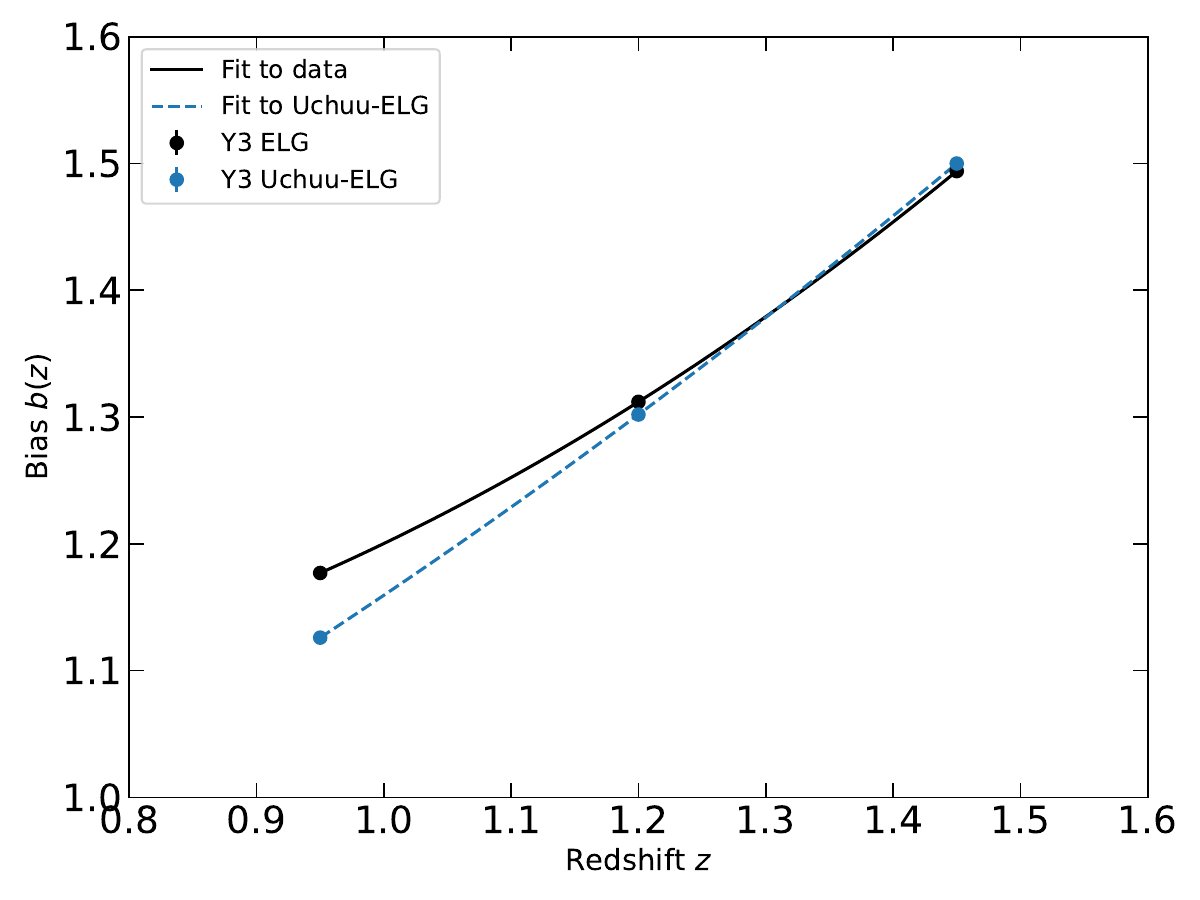}
    \hspace{0.02\textwidth}
    \includegraphics[width=0.49\columnwidth]{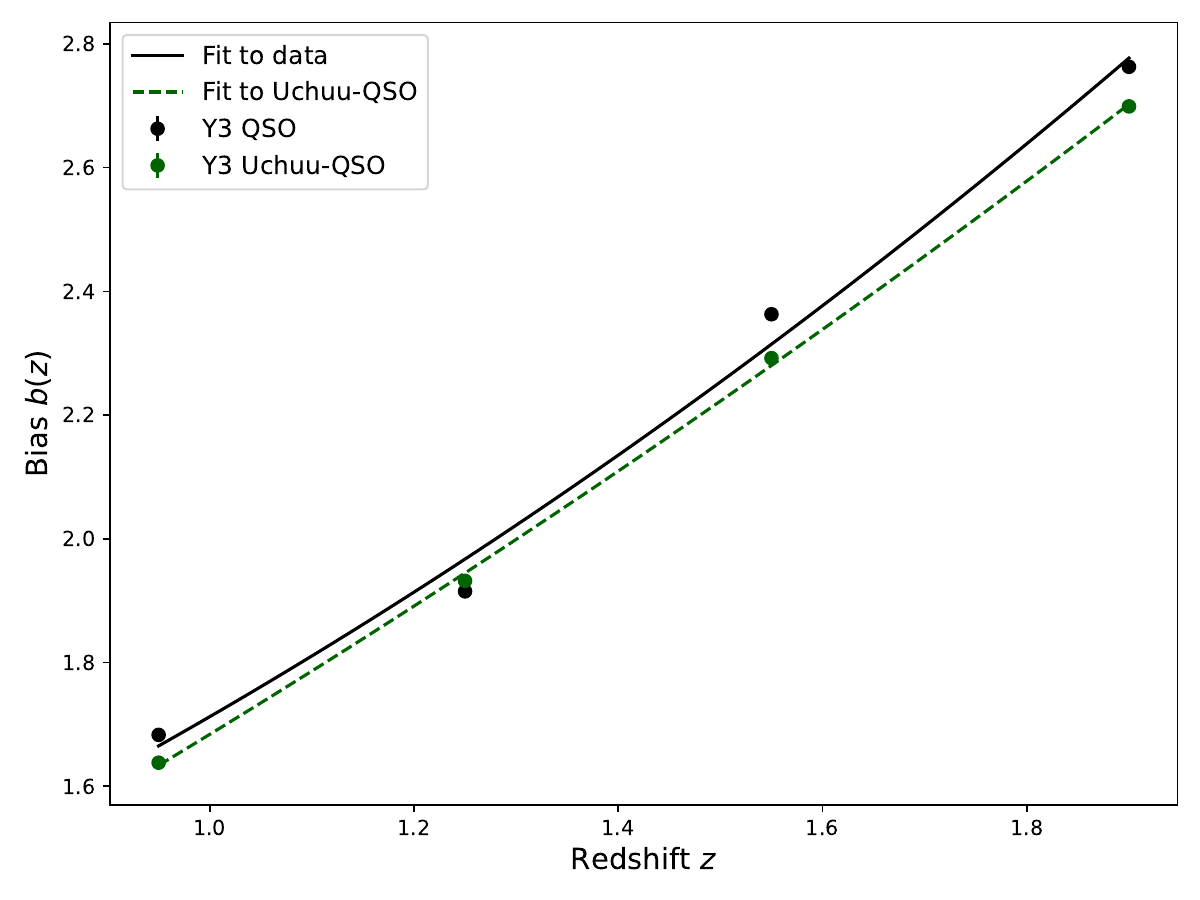}
    \caption{Large-scale bias $b(z)$ as a function of redshift for ELGs (left) and QSOs (right). 
Black and colored points show bias values measured directly from DESI Y3 observations 
and Uchuu mocks, respectively, with solid and dashed lines representing their 
corresponding fits (as mentioned in \ref{subsec:bias}). The bias increases monotonically with redshift for both 
tracers, consistent with these populations occupying more massive halos at earlier 
times.}
    \label{fig:bias}
\end{figure*}

As we discussed in \ref{sec:intro}, we measure the large-scale bias, $b$, for both tracers from the DESI DR2 and compare to their prediction obtained from the \Uchuu lightcones in the Planck cosmology. The results are presented in Figure~\ref{fig:bias}. We performed these measurements of the linear bias by fitting 
\begin{equation}
    \xi_0(s) = b^2 \left(1 + \frac{2}{3}\beta + \frac{1}{5}\beta^2 \right) \xi_\mathrm{lin}(s)
\end{equation}
to our correlation function monopole measurements, $\xi_0(s)$, over a given range of separations. $\xi_\mathrm{lin}(s)$ is from the linear power spectrum at the redshift of our galaxy sample, and $\beta=\Omega_\mathrm{m}^{0.6}/b$ \citep[see][]{Kaiser1987,Hamilton1998}. 

The bias of ELGs vs redshift is shown in the left panel of Figure~\ref{fig:bias}. The bias was calculated for each box defined in Table~\ref{tab:elg-shamparm} and for the data cut into the same redshift range covered by the mocks. We use a separation range of $10~\hMpc$ to $80~\hMpc$. We find good agreement between the bias measured from the data and the bias measured from the mock. 
\noindent
The large-scale bias of the QSOs is shown in the right panel of Figure~\ref{fig:bias}, as a function of redshift, with bias factor measured in the separation range $10 < s < 80~\hMpc$. The increasing bias measurements with redshift are consistent with those obtained by \citep{KrolewskiQSObias} using the DESI two-month data. 

\section{Summary}
\label{sec:concl}

The work presented in this paper provides a detailed overview of the work to create simulated galaxy catalogs in lightcones for the DESI DR2. These lightcones are constructed within the framework of the flat-$\Lambda$CDM Planck cosmology model implemented in the \Uchuu cosmological simulation. We construct catalogs populated by DESI ELGs and QSOs with a general model of $V_\mathrm{peak}$ of galaxies in (sub)halos and abundance match to (sub)halo mass. A modified SHAM method is employed, incorporating additional parameters (for ELGs). Our models, that account for a $V_\mathrm{peak}$ (for QSOs) or $f_\mathrm{sat}$ evolving with redshift, predict the observations reasonably across different narrow redshift bins.
We carry out a comparison of the measured clustering signals for ELG and QSO sample in the DESI DR2 with the corresponding predictions from \Uchuu. Additionally, we determine the halo occupancy and large-scale bias factors for both tracers. We summarize the main results below:

\begin{enumerate}
    \item We measure the redshift-space two-point correlation function monopole and quadrupole over the scales $0.5~\hMpc$ to $80~\hMpc$. Additionally, we measure the power spectrum monopole from $0.005 ~h\mathrm{Mpc}^{-1}< k < 0.405~h\mathrm{Mpc}^{-1}$.
    Overall, we find consistency between the DR2 measurements and the theoretical predictions based on the Planck cosmology using the \Uchuu lightcones.

    \item For the ELGs, there is agreement between \Uchuu and DESI above $0.5~\hMpc$.
    \item For QSOs, good agreement is found between \Uchuu and the DESI DR2 data in the respective redshift bins.

    \item The ELG halo occupation consists of a Gaussian component for centrals with low halo masses dominating at ($M_{\rm halo} < 10^{12.1}~\hMsun$ and a peak occupation of 0.095 at $M_{\rm halo} = 10^{12}~\hMsun$), and a power law component for satellites with halo masses dominating at ($M_{\rm halo} > \times10^{12.4}~\hMsun$). The QSO HOD exhibits a similar form: a Gaussian component for central haloes dominating at ($M_{\rm halo} < 10^{12.7}~\hMsun$ with a peak occupation of 0.319 at $M_{\rm halo} = 10^{12.4}~\hMsun$), and a power-law component for the satellites dominating at ($M_{\rm halo} > 10^{12.8}~\hMsun$).
    \item The linear bias factors were measured for both tracers from the DESI DR2 and compared to predictions based on the \Uchuu lightcones in the Planck cosmology. 
    
\end{enumerate} 



\section*{Acknowledgements}

RK acknowledges support of the U.S. Department of Energy (DOE) in funding grant DE-SC0010129 for the work in this paper.
Computational resources for RV, JL, RK, AA, and NK were provided by SMU's Center for Research Computing.
The \Uchuu simulation was carried out on the Aterui II supercomputer at CfCA-NAOJ. We thank IAA-CSIC, CESGA, and RedIRIS in Spain for hosting the Uchuu data releases in the \textsc{Skies \& Universes} site for cosmological simulations. The analysis performed for this paper have employed NERSC at LBNL and $skun6$@IAA-CSIC computer facility managed by IAA-CSIC in Spain (MICINN EU-Feder grant EQC2018-004366-P).

This material is based upon work supported by the U.S. Department of Energy (DOE), Office of Science, Office of High-Energy Physics, under Contract No. DE–AC02–05CH11231, and by the National Energy Research Scientific Computing Center, a DOE Office of Science User Facility under the same contract. Additional support for DESI was provided by the U.S. National Science Foundation (NSF), Division of Astronomical Sciences under Contract No. AST-0950945 to the NSF’s National Optical-Infrared Astronomy Research Laboratory; the Science and Technology Facilities Council of the United Kingdom; the Gordon and Betty Moore Foundation; the Heising-Simons Foundation; the French Alternative Energies and Atomic Energy Commission (CEA); the Secretariat of Science, Humanities, Technology and Innovation (SECIHTI) of Mexico; the Ministry of Science, Innovation and Universities of Spain (MICIU/AEI/10.13039/501100011033), and by the DESI Member Institutions: \url{https://www.desi.lbl.gov/collaborating-institutions}. Any opinions, findings, and conclusions or recommendations expressed in this material are those of the author(s) and do not necessarily reflect the views of the U. S. National Science Foundation, the U. S. Department of Energy, or any of the listed funding agencies.

The authors are honored to be permitted to conduct scientific research on I'oligam Du'ag (Kitt Peak), a mountain with particular significance to the Tohono O’odham Nation.

\section*{Data Availability}
The data
points corresponding to the figures from this paper will be available in the Zenodo repository at \url{https://doi.org/10.5281/zenodo.20907446} .



\appendix

\section*{Appendix}
\addcontentsline{toc}{section}{Appendix}

\section{Example of $V_{\rm peak}$ distribution}
\label{app:Vpeak_dist}

This section provides an example of the $V_{\rm peak}$ distribution for one of the QSO mock catalogs used in the parameter grid. The purpose of this figure is to illustrate that the range and shape of the halo maximum circular velocity distribution selected by the model, (the adopted $V_{\rm peak}$ selection) contributes to the resulting clustering signal.

\begin{figure}[H]
    \centering
    \includegraphics[width=0.45\linewidth]{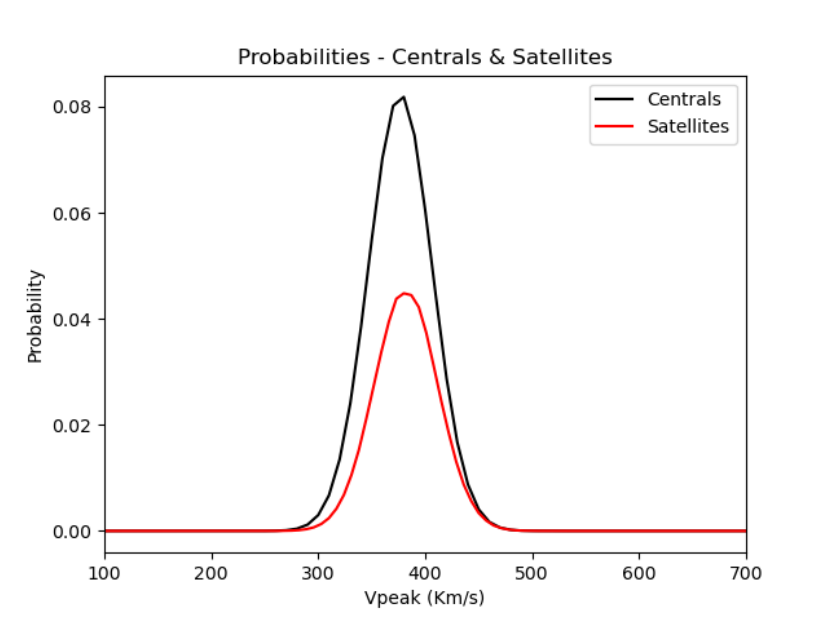}
    \caption{Distribution of $V_{\rm peak}$ for one of the QSO mocks from the grid.}
    \label{fig:vpeakhist}
\end{figure}

\section{Effect of Sigma variation}
\label{app:sigma-variation}

This section examines how changing the velocity scatter parameter $\sigma_v$ affects the predicted clustering of the mock samples. Across the fitting scales considered in this work, the clustering measurements remain largely insensitive to variations in $\sigma_v$, indicating that this parameter has a negligent impact compared with parameters such as the mean $V_{\rm peak}$ and satellite fraction. It is worth noting that this maybe explored further for future studies investigating small scale clustering. 
\begin{figure}[H]
    \centering
    \includegraphics[width=0.45\linewidth]{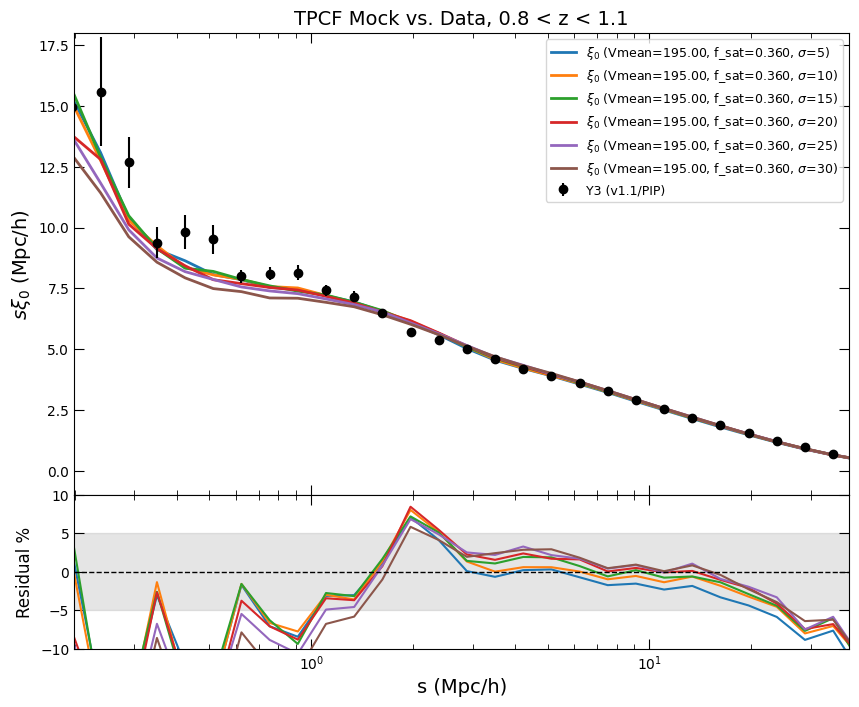}
    \caption{Effect of $\sigma_v$ variation in ELG mocks ($0.8 < z < 1.1$)}
    \label{fig:sigmavary}
\end{figure}

\section{Heatmaps for best fit parameters}
\label{app:heatmaps-best-fit}

This section presents a heatmap of the reduced $\chi^2$ values obtained across the model parameter grid as described in Section~\Ref{sec:uchuu_mod_sham}, allowing the preferred regions of parameter space to be identified visually. In these plots, each point corresponds to a mock realization with a particular combination of mean $V_{\rm peak}$ and satellite fraction $f_{\rm sat}$, and the color scale indicates the quality of fit to the observed clustering. The heatmap therefore provides a compact summary of the parameter degeneracies and highlights the combinations that best reproduce the data, helping to justify the final best-fit parameter choices reported in the main analysis.

\begin{figure}
    \centering
    \includegraphics[width=0.65\linewidth]{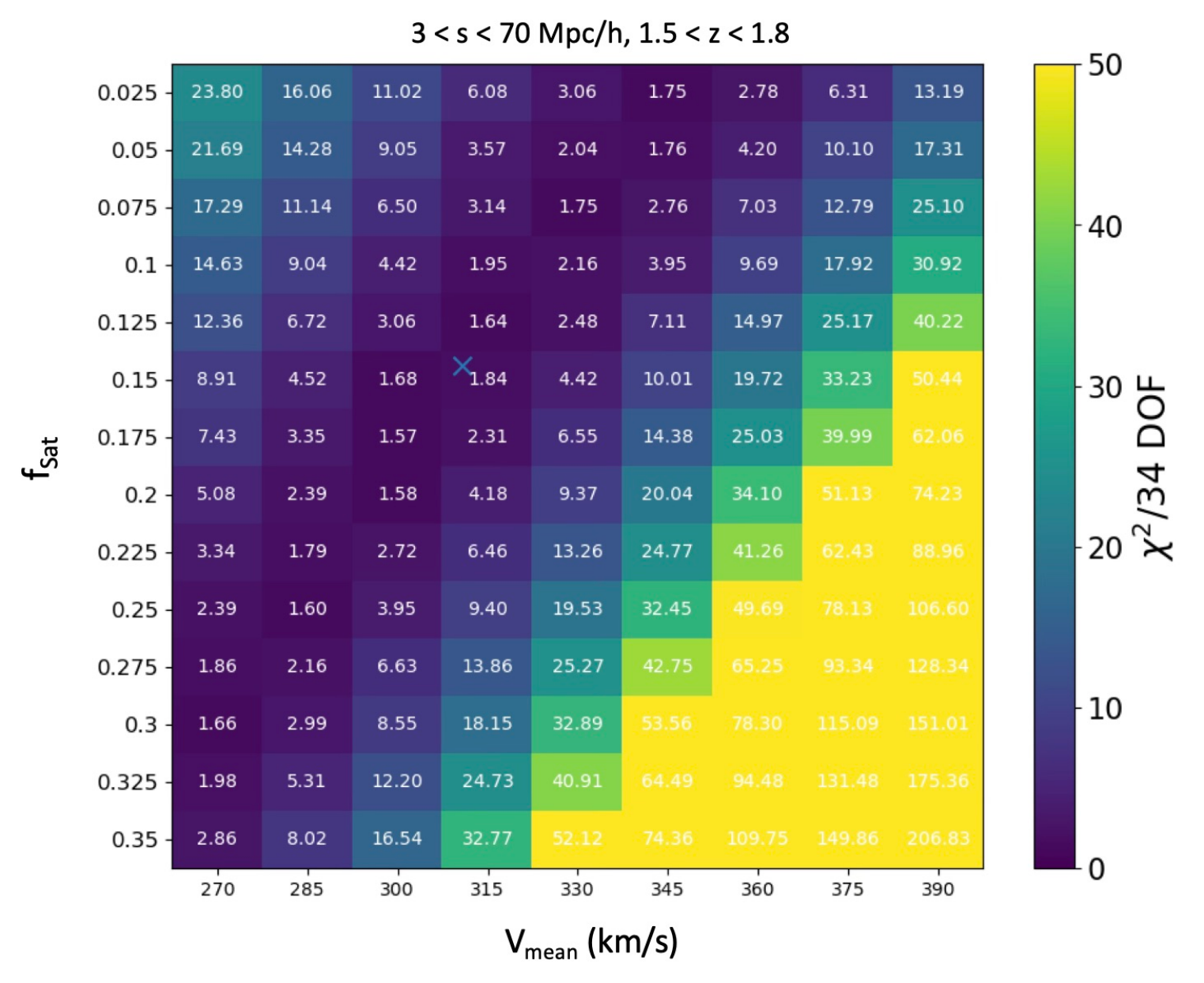}
    \caption{Heatmap for $\chi^2$ per degrees of freedom (reduced $\chi^2$) for QSOs in $1.5 < z < 1.8$. The x-axis is the mean $V_{\rm peak}$ and the y-axis is the respective satellite fractions $f_{\rm sat}$. The blue cross indicates the region of parameter space chosen by the the paraboloid fit as the best fir region for this redshift region. }
    \label{fig:placeholder}
\end{figure}

\section{Affiliations}
\label{app:affil}

\begin{hangparas}{.5cm}{1}

$^{a}${Department of Physics, Southern Methodist University, 3215 Daniel Avenue, Dallas, TX 75275, USA}

$^{b}${Astrophysics \& Space Institute, Schmidt Sciences, New York, NY 10011, USA}

$^{c}${Instituto de Astrof\'{i}sica de Andaluc\'{i}a (CSIC), Glorieta de la Astronom\'{i}a, s/n, E-18008 Granada, Spain}

$^{d}${Institute for Astronomy, University of Edinburgh, Royal Observatory, Blackford Hill, Edinburgh EH9 3HJ, UK}

$^{e}${Physics Department, Brookhaven National Laboratory, Upton, NY 11973, USA}

$^{f}${Lawrence Berkeley National Laboratory, 1 Cyclotron Road, Berkeley, CA 94720, USA}

$^{g}${Center for Cosmology and AstroParticle Physics, The Ohio State University, 191 West Woodruff Avenue, Columbus, OH 43210, USA}

$^{h}${Department of Astronomy, The Ohio State University, 4055 McPherson Laboratory, 140 W 18th Avenue, Columbus, OH 43210, USA}

$^{i}${The Ohio State University, Columbus, 43210 OH, USA}

$^{j}${Department of Physics, Boston University, 590 Commonwealth Avenue, Boston, MA 02215 USA}

$^{k}${Dipartimento di Fisica ``Aldo Pontremoli'', Universit\`a degli Studi di Milano, Via Celoria 16, I-20133 Milano, Italy}

$^{l}${INAF-Osservatorio Astronomico di Brera, Via Brera 28, 20122 Milano, Italy}

$^{m}${Department of Physics \& Astronomy, University College London, Gower Street, London, WC1E 6BT, UK}

$^{n}${Institut d'Estudis Espacials de Catalunya (IEEC), c/ Esteve Terradas 1, Edifici RDIT, Campus PMT-UPC, 08860 Castelldefels, Spain}

$^{o}${Institute of Space Sciences, ICE-CSIC, Campus UAB, Carrer de Can Magrans s/n, 08913 Bellaterra, Barcelona, Spain}

$^{p}${Department of Physics and Astronomy, The University of Utah, 115 South 1400 East, Salt Lake City, UT 84112, USA}

$^{q}${Instituto de F\'{\i}sica, Universidad Nacional Aut\'{o}noma de M\'{e}xico,  Circuito de la Investigaci\'{o}n Cient\'{\i}fica, Ciudad Universitaria, Cd. de M\'{e}xico  C.~P.~04510,  M\'{e}xico}

$^{r}${University of California, Berkeley, 110 Sproul Hall \#5800 Berkeley, CA 94720, USA}

$^{s}${Departamento de F\'isica, Universidad de los Andes, Cra. 1 No. 18A-10, Edificio Ip, CP 111711, Bogot\'a, Colombia}

$^{t}${Observatorio Astron\'omico, Universidad de los Andes, Cra. 1 No. 18A-10, Edificio H, CP 111711 Bogot\'a, Colombia}

$^{u}${Institute of Cosmology and Gravitation, University of Portsmouth, Dennis Sciama Building, Portsmouth, PO1 3FX, UK}

$^{v}${University of Virginia, Department of Astronomy, Charlottesville, VA 22904, USA}

$^{w}${Fermi National Accelerator Laboratory, PO Box 500, Batavia, IL 60510, USA}

$^{x}${Department of Astronomy, University of Texas at Austin, 2515 Speedway, TX 78712, USA}

$^{y}${Department of Physics, The University of Texas at Dallas, 800 W. Campbell Rd., Richardson, TX 75080, USA}

$^{z}${NSF NOIRLab, 950 N. Cherry Ave., Tucson, AZ 85719, USA}

$^{aa}${Departament de F\'{i}sica, Serra H\'{u}nter, Universitat Aut\`{o}noma de Barcelona, 08193 Bellaterra (Barcelona), Spain}

$^{ab}${Institut de F\'{i}sica d’Altes Energies (IFAE), The Barcelona Institute of Science and Technology, Edifici Cn, Campus UAB, 08193, Bellaterra (Barcelona), Spain}

$^{ac}${Instituci\'{o} Catalana de Recerca i Estudis Avan\c{c}ats, Passeig de Llu\'{\i}s Companys, 23, 08010 Barcelona, Spain}

$^{ad}${Department of Physics \& Astronomy, University  of Wyoming, 1000 E. University, Dept.~3905, Laramie, WY 82071, USA}

$^{ae}${Department of Physics and Astronomy, University of Waterloo, 200 University Ave W, Waterloo, ON N2L 3G1, Canada}

$^{af}${Perimeter Institute for Theoretical Physics, 31 Caroline St. North, Waterloo, ON N2L 2Y5, Canada}

$^{ag}${Waterloo Centre for Astrophysics, University of Waterloo, 200 University Ave W, Waterloo, ON N2L 3G1, Canada}

$^{ah}${Departament de F\'isica, EEBE, Universitat Polit\`ecnica de Catalunya, c/Eduard Maristany 10, 08930 Barcelona, Spain}

$^{ai}${Department of Physics and Astronomy, Sejong University, 209 Neungdong-ro, Gwangjin-gu, Seoul 05006, Republic of Korea}

$^{aj}${CIEMAT, Avenida Complutense 40, E-28040 Madrid, Spain}

$^{ak}${Department of Physics \& Astronomy, Ohio University, 139 University Terrace, Athens, OH 45701, USA}

$^{al}${University of Michigan, 500 S. State Street, Ann Arbor, MI 48109, USA}

$^{am}${National Astronomical Observatories, Chinese Academy of Sciences, A20 Datun Road, Chaoyang District, Beijing, 100101, P.~R.~China}

\end{hangparas}

\newpage



\bibliographystyle{JHEP}
\bibliography{biblio}


\label{lastpage}
\end{document}